\documentclass[aps,prc,floatfix,showpacs,twocolumn]{revtex4-1}
\usepackage{dcolumn}
\usepackage{bm}
\usepackage{color}
\usepackage{amssymb}
\usepackage{amsmath}
\usepackage{graphicx}
\usepackage{amsfonts}
\usepackage{slashed}
\usepackage{pstricks}
\usepackage{float}
\usepackage{hyperref}
\usepackage{array}
\allowdisplaybreaks
\usepackage{dcolumn}
\usepackage{epsf}
\newcommand{\nmax}{N_{\mbox{\footnotesize{max}}}}

\begin{document}
\title{Inclusive electron-nucleus cross section within the Self Consistent \\ Green's Function approach}
\author{
N.\ Rocco$^{\, {\rm a} }$,
C.\ Barbieri$^{\, {\rm a} }$,
}
\affiliation{
$^{\,{\rm a}}$\mbox{Department of Physics, University of Surrey, Guildford, GU2 7HX, UK}\\
}
\date{\today}

\date{\today}
\begin{abstract}
We compute inclusive electron-nucleus cross sections using \textit{ab initio} spectral functions of $^4$He and $^{16}$O obtained within the Self Consistent Green's Function approach. The formalism adopted is based on the factorization of the spectral function and the nuclear transition matrix elements. This allows to provide an accurate description of nuclear dynamics and to account for relativistic effects in the interaction vertex.  Our calculations use a saturating chiral Hamiltonian in order reproduce the correct nuclear sizes.  When final state interactions for the struck particle are accounted for, we find nice agreement between the data and the theory for the inclusive electron-$^{16}$O cross section.  The results lay the foundations for future applications of the Self Consistent Green's Function method, in both closed and open shell nuclei, to neutrino data analysis. 

 This work also presents results for the point-proton, charge and single-nucleon momentum distribution of the same two nuclei. 
 The center of mass can affect these quantities for light nuclei and cannot be separated cleanly in most \textit{ab initio} post-Hartree-Fock  methods. 
In order to address this, we developed a Metropolis Monte Carlo calculation in which the center of mass coordinate can be subtracted  exactly from the trial wave function and the expectation values. 
 We gauged this effect for  $^4$He by removing the center of mass effect  from the  Optimal Reference State wave function that is generated during the Self Consistent Green's Function calculations.
 Our findings clearly indicate that the residual center of mass contribution strongly modifies calculated matter distributions with
respect to those obtained in the intrinsic frame. Hence, its subtraction is crucial for a correct description of light nuclei.
\end{abstract}
\maketitle

\section{Introduction}
The current- and next-generation of neutrino oscillation experiments require nuclear physics calculations of the structure and electroweak properties of atomic nuclei supplemented by quantified 
theoretical uncertainties~\cite{Gran:2006jn,AguilarArevalo:2007ab,Lyubushkin:2008pe}. The Deep Underground Neutrino Experiment (DUNE) will exploit Liquid Argon Time-Projection Chambers (TPCs) to test CP violation in the lepton sector
and to shed light on the neutrino mass hierarchy.  Hence, nuclear theories 
able to tackle genuine open shell nuclei, such as Argon, will be critical to the reconstruction of the initial neutrino energy.

The Self Consistent Green's Function (SCGF) approach is an \textit{ab initio} method in which the
optical potential and spectral functions are calculated covering the full spectra of both nucleon attachment and removal (\textit{i.e.}, both close and far form the Fermi surface)~\cite{Dickhoff:2004xx,Barbieri2003DRPA,Barbieri2006plbO16,Barbieri:2007zy,Barbieri:2009nx,Cipollone:2013zma}. 
The  self-consistency feature means that the input information about the ground state and 
excitations of the systems no longer depends on a user-defined reference state but instead it is taken directly from the computed 
correlated propagator. 
The SCGF method has recently been reformulated within Gorkov's theory that allows to address open shell nuclei~\cite{Soma:2011aj,Soma:2012zd,Soma:2013ona}. Within this approach, the description of pairing 
correlations characterizing open shell systems is achieved by breaking particle number symmetry.
The method was extended to include three-body interactions in Ref.~\cite{Carbone:2013eqa,Carbone2014corr3nf}.
Modern two- and three-nucleon chiral forces can be fully exploited within this formalism. 
Because of these features, SCGF theory is a prime tool for providing the nuclear structure input necessary to calculate electroweak properties of nuclei. However, its performance in predicting lepton-nucleus reactions with chiral nuclear forces is still to be assessed.

In this work we use the saturating next-to next-to leading order (NNLO) interaction denoted as NNLO$_{\rm{sat}}$~\cite{Ekstrom:2015rta} and calculate the SCGF spectral functions of $^{4}$He and $^{16}$O. 
We obtain their point density, charge density and single-momentum distribution. All calculations are performed expanding on an harmonic oscillator basis and the dependence of the results on the oscillator parameters is investigated.
For light nuclei, such as $^4$He, spurious contributions of the center of mass in the calculated wave functions can be sizable in the model spaces exploited by SCGF and other post-Hartree-Fock methods. While it is possible to show that the center of mass effectively decouples from the relative motion for large enough spaces~\cite{Hagen2009PRLcom},  subtracting its effect from the calculated wave function and spectral functions is a nontrivial long standing issue.
Here, we address this problem by performing a Monte Carlo integration in which the center of mass component is exactly subtracted from the  wave function. 

In the high momentum transfer region of neutrino-nucleus scattering,  the formalism based on spectral function and 
factorization of the nuclear transition matrix elements allows to combine a fully relativistic description of the single-nucleon 
interaction vertex with an accurate treatment of nuclear dynamics~\cite{Benhar:2005dj,Benhar:2006wy}.
In order to apply any theoretical model in the neutrino data analysis, it is fundamental to validate it against the large body of 
electron scattering data. This work has to be considered as a first step in this direction. In fact, we present 
an extensive comparison with the experimental data of the electromagnetic double differential cross sections of
$^4$He and $^{16}$O scattering, which was obtained by exploiting the corresponding SCGF spectral functions.
 The predictions for $^{16}$O are important for the data analysis of Super-Kamiokande, in which a water Cherenkov
 detector to study neutrinos produced from different sources is used.

In Section \ref{sec:SCGF} we briefly review the SCGF formalism and the links of propagators with the one-body density
distribution and other quantities of experimental interest.
Section \ref{IA:appr} is devoted to the derivation of the electron-nucleus cross section
within the Impulse Approximation (IA) in which the factorization of the nuclear transition matrix elements
is assumed. Final state interactions (FSI) involving the struck particle are treated as corrections. They are included using the convolution approach of Refs.~\cite{Ankowski:2014yfa,Benhar:1991af}.
In Section \ref{results} we present results for
the point density, charge density and single-momentum distribution of $^4$He and $^{16}$O.
In addition, the inclusive electromagnetic cross sections of these two nuclei, obtained
using the associated SCGF spectral functions, are compared with the experimental data and the role played by
FSI is discussed.
Conclusions are drawn in Section~\ref{concl}.

\section{The Self Consistent Green's Function Approach}
\label{sec:SCGF}
\begin{figure}[H]
\centering
\includegraphics[scale=0.300]{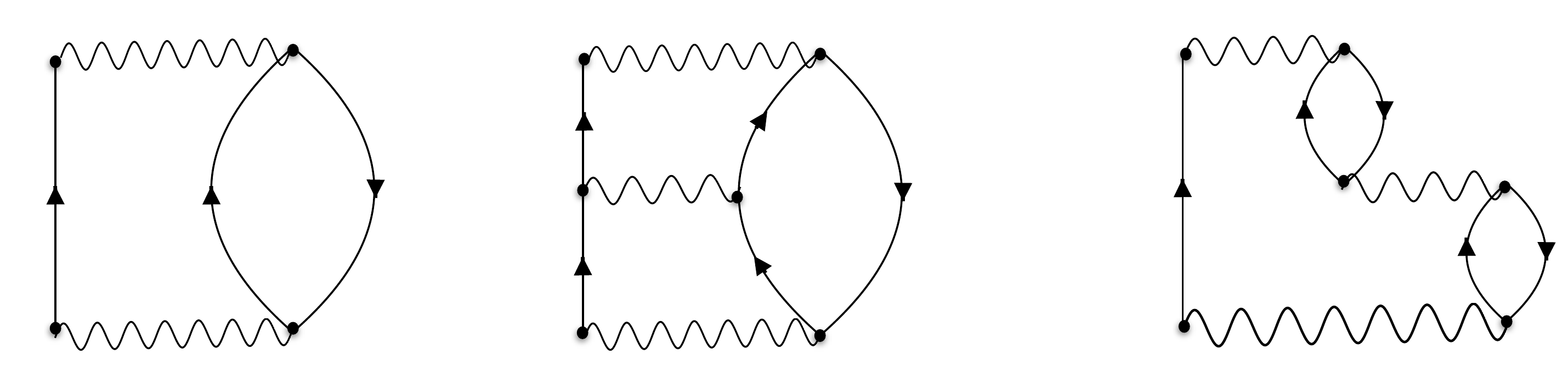}
\caption{One-particle irreducible skeleton and interaction irreducible diagrams with $2h1p$ and $2p1h$ intermediate configurations. 
The wiggly lines represent the two-body effective interaction of Eq.~\eqref{12b:eff:int}.}
\label{2h1p:diag}
\end{figure}

The one-body Green's Function is written as a sum of two different contributions describing the propagation 
of a particle and hole state~\cite{Dickhoff2008book}:
\begin{align}
g_{\alpha\beta}(\omega)=&\langle \psi_0^A| a_\alpha \frac{1}{\omega-(H-E^A_0)+i\eta} a^\dagger_\beta |\psi_0^A\rangle\nonumber\\
&+\langle \psi_0^A| a^\dagger_\beta \frac{1}{\omega+(H-E^A_0)-i\eta} a_\alpha |\psi_0^A\rangle\, ,
\label{gf:def}
\end{align}
where $a^\dagger_\alpha$ and $a_\alpha$ are the creation and annihilation operator in the quantum 
state $\alpha$, respectively.
The so-called Lehmann representation results from inserting completeness relations in Eq.~\eqref{gf:def}. This is
\begin{align}
g_{\alpha\beta}(\omega)=&\sum_n\frac{\langle \psi_0^A| a_\alpha |\psi_n^{A+1}\rangle \langle\psi_n^{A+1}|a^\dagger_\beta|\psi_0^A\rangle}{\omega -(E_n^{A+1}-E_0^A)+i\eta}\nonumber\\
& \sum_k \frac{\langle \psi_0^A| a^\dagger_\beta |\psi_k^{A-1}\rangle \langle\psi_k^{A-1}|a_\alpha|\psi_0^A\rangle}{\omega -(E_0^{A}-E_k^{A-1})-i\eta}\, ,
\label{lehm:dec}
\end{align}
where $|\psi_n^{A+1}\rangle$ ($|\psi_k^{A-1}\rangle$) are the eigenstates and $E_n^{A+1}$ ($E_k^{A-1}$)  the eigenvalues of the $(A\pm1)$-body system.
Introducing the transition amplitudes
\begin{align}
({\mathcal X}^n_\alpha)^\ast={}&\langle \psi_0^A| a_\alpha |\psi_n^{A+1}\rangle\, ,\nonumber\\
{\mathcal Y}^k_\alpha={}& \langle\psi_k^{A-1}|a_\alpha|\psi_0^A\rangle
\label{tr:ampl}
\end{align}
and the corresponding quasiparticle energies
\begin{align}
\epsilon^+_n={}&E_n^{A+1}-E_0^A  \, ,\nonumber\\
\epsilon^-_k={}&E_0^{A}-E_k^{A-1}  
\label{tr:qpens}
\end{align}
leads to the more compact expression
\begin{align}
g_{\alpha\beta}(\omega)={}&\sum_n\frac{({\mathcal X}^n_\alpha)^\ast \, {\mathcal X}^n_\beta}{\omega -\epsilon^+_n+i\eta}
~+~
 \sum_k \frac{ {\mathcal Y}^k_\alpha \,  ({\mathcal Y}^k_\beta)^\ast }{\omega -\epsilon^-_k-i\eta}\, .
\label{corr:prop}
\end{align}
The one-body propagator given in Eqs.~\eqref{gf:def} and ~\eqref{lehm:dec} is completely determined by solving the Dyson equation
\begin{align}
g_{\alpha\beta}(\omega)=g^0_{\alpha\beta}(\omega)+\sum_{\gamma\delta}g^0_{\alpha\gamma}(\omega)\Sigma^\star_{\gamma\delta}(\omega)g_{\delta\beta}(\omega) \, ,
\end{align}
where $g^0_{\alpha\beta}(\omega)$ is the unperturbed single-particle propagator and $\Sigma^\star_{\gamma\delta}(\omega)$ is the irreducible self-energy that encodes nuclear medium effects in the particle propagator~\cite{Dickhoff2008book}. 
The latter is given by the sum of two different terms
\begin{align}
\Sigma^\star_{\alpha\beta}(\omega)=\Sigma^\infty_{\alpha\beta}+\tilde{\Sigma}_{\alpha\beta}(\omega)\ ,
\label{self:en}
\end{align}
the first one describes the average mean field while the second one contains dynamic correlations.
In practical calculations the self-energy is expanded as a function of the propagator itself, implying that
an iterative procedure is required to solve the Dyson equation self-consistently. 
The self-energy can be calculated systematically within the Algebraic Diagrammatic Construction (ADC) method. The third order truncation of this
scheme [ADC(3)] yields a propagator that includes  all possible Feynman contributions up to third order but it further resums infinite series of relevant diagrams in a non-perturbative fashion~\cite{Schirmer1983,Barbieri2017LNP}.
Two- and three-nucleon force contributions are included.
A first organization of the contributions to the self-energy comes by considering the particle irreducible (PI) and skeleton diagrams.
The number of Feynman diagrams entering the calculation of the Green's Function rapidly increase when three- or many-body forces 
are accounted for. 
In order to circumvent this problem and reduce the number of Feynman diagrams to be considered, a useful strategy is to 
include only interaction-irreducible diagrams~\cite{Carbone:2013eqa}. For our calculations, we use the following medium dependent or effective one- and two-body interactions:
\begin{align}
&\tilde{U}_{\alpha\beta}=U_{\alpha\beta}+\sum_{\delta\gamma}V_{\alpha\gamma,\beta\delta}\rho_{\delta\gamma}+\frac{1}{4}\sum_{\mu\nu\gamma\delta}W_{\alpha\mu\nu,\beta\gamma\delta}
\rho_{\gamma\mu}\rho_{\nu\delta}\, ,\nonumber\\
&\tilde{V}_{\alpha\beta,\delta\gamma}=V_{\alpha\beta,\delta\gamma}+\sum_{\mu\nu}W_{\alpha\beta\mu,\gamma\delta\nu}\rho_{\nu\mu}\, ,
\label{12b:eff:int}
\end{align}
where $U,V$ and, $W$ label the matrix elements of the \hbox{one-}, two-, and three-body interactions, respectively. 
The one-body  density matrix appearing in Eq.~\eqref{12b:eff:int} reads
\begin{align}
\label{1b:dens}
&\rho_{\delta\gamma}=\langle \psi_0^A|a^\dagger_\gamma a_\delta|\psi_0^A\rangle\, .
\end{align}
The use of this averaging procedure allows to retain only interaction irreducible diagrams in the effective interactions $\tilde{U}$ and $\tilde{V}$, 
while residual contributions that include  $W$ can be safely neglected~\cite{Hagen:2007ew,Roth:2011vt,Barbieri2014QMBT,Cipollone:2014hfa}.
 The expressions of the static and dynamic self-energy up to third order, including all possible two- and three-nucleon terms that enter the expansion of the self-energy,
 as well as interaction-irreducible (\textit{i.e.} not averaged) three-nucleon diagrams have been recently derived in Ref.~\cite{Raimondi:2017kzi}.  \\
Figure~\ref{2h1p:diag} displays the three simplest diagrams that enter the present calculation of the self-energy. These
are taken as ``seeds'' for an all order resummation that eventually generates $\Sigma^\star_{\alpha\beta}(\omega)$.
The first contribution is at second-order while the last two are of third-order in the expansion of Eq.~\eqref{self:en}. 
Note that, for all the considered diagrams, the set of intermediate state configurations corresponds to two-particles--one-hole (2p1h) and 
two-holes--one-particle (2h1p) and that we use the two-nucleon effective interaction of Eq.~\eqref{12b:eff:int} .
Within the ADC(3) approach an infinite order summation of diagrams of Fig.~\ref{2h1p:diag}  that includes 
particle-particle and hole-hole ladders as well as particle-hole rings is performed.
The dynamical part of the self-energy of Eq.\eqref{self:en} can be rewritten in the Lehmann representation as
\begin{align}
\tilde{\Sigma}_{\alpha\beta}(\omega)=\sum_{ij^\prime} {\bf D}^{\dagger}_{\alpha i} \Big[\frac{1}{\omega-({\bf K}+{\bf C})} \Big]_{ij}{\bf D}^{\dagger}_{j \beta}\ ,
\label{self:en2}
\end{align}
where ${\bf K}$ are the unperturbed 2p1h and 2h1p energies, ${\bf D}$ coupling matrices and {\bf C} interaction matrices for the forward and backward intermediate states.

The calculations presented in this work have been performed expanding one-, two- and three-body operators on a spherical harmonic oscillator basis whose dimension and the
oscillation frequency are denoted by $\nmax=\max\{2n+\ell\}$ and $\hbar \Omega$, respectively.

The point-proton density distribution can be readily obtained from Eq.\eqref{1b:dens} and reads
\begin{align}
\rho_p({\bf r})=\sum_{\alpha\beta}\Phi^\ast_\beta({\bf r})\Phi_\alpha({\bf r})\rho_{\alpha\beta} \, ,
\end{align}
where $\Phi_\alpha({\bf r})=\langle {\bf r} |\alpha\rangle$  denotes the harmonic oscillator singe-particle wave-functions and the sum includes only proton single-particle states. Analogous expressions can be written for point-neutron and matter density distributions.

The computational cost required to account for the fragmentation of the single-particle propagator into Eq.~\eqref{self:en2}
rapidly increases with the size of the nucleus and of the model space. 
For this reason an Optimized Reference State (OpRS) approach is used to approximate the single particle propagators entering in the 
diagrams of Fig.~\ref{2h1p:diag}~\cite{Barbieri:2009nx}.
The OpRS is taken to be an independent particle model propagator as
\begin{align}
g_{\alpha\beta}^{\rm{OpRS}}=\sum_{n \not\in F}\frac{(\phi^n_\alpha)^\ast\phi^n_\beta}{\omega-\epsilon^{\rm{OpRS}}_n+i\eta}+\sum_{k \in F}\frac{\phi^k_\alpha(\phi^k_\beta)^\ast}{\omega-\epsilon^{\rm{OpRS}}_k-i\eta}
\end{align}
where $F$ represents the set of occupied states, $\epsilon^{\rm{OpRS}}$ and $\phi$ are the single particle energies and wave functions, respectively.
The OpRS propagator is chosen to best approximate the correlated one while keeping a reduced number of poles. 
This is achieved by introducing the following momenta of the spectral distribution with respect to energy poles:
\begin{align}
M^0_{\alpha\beta}=&\sum_n({\cal X}^n_\alpha)^\ast {\cal X}^n_\beta +\sum_k {\cal Y}^k_\alpha ({\cal Y}^k_\beta)^\ast  \, ,\nonumber\\
M^1_{\alpha\beta}=&\sum_n\frac{({\cal X}^n_\alpha)^\ast {\cal X}^n_\beta}{(E_F-\epsilon^+_n)} +\sum_k \frac{{\cal Y}^k_\alpha ({\cal Y}^k_\beta)^\ast}{(E_F-\epsilon^-_k)} \, ,
\label{eq:M0M1}
\end{align}
where $E_F=(\epsilon^+_0+\epsilon^-_0)/2=(E_0^{A-1}+E_0^{A+1})/2$ and the transition amplitudes are given in Eqs.~\eqref{tr:ampl}.
The quantities in Eq.~\eqref{eq:M0M1} are important since they constrain the density distributions, one-body observables and the Koltun energy sum rule of the propagator~\cite{Rios2017SumRules}. 
Hence, we obtain $\epsilon^{\rm{OpRS}}$ and $\phi$ by requiring that the OpRS lowest momenta of the spectral distribution reproduce those of the full calculation, \textit{i.e.}
 $M^{0,OpRS}_{\alpha\beta}=M^0_{\alpha\beta}$ and $M^{1,OpRS}_{\alpha\beta}=M^1_{\alpha\beta}$~\cite{Barbieri:2009nx}.\\

The elastic scattering of a nucleus hit by a probe and recoiling with a momentum
${\bf q}$ is described by the elastic form factor $F_L({\bf q})$.
Neglecting the small spin-orbit contribution, the latter is given by
\begin{align}
F_L({\bf q})=&\frac{1}{Z}\langle \psi_0^A|\Big[\frac{\sum_i\epsilon_i}{\sqrt{1+Q^2_{el}/(4m^2)}} \Big]|\psi_0^A\rangle\, ,
\end{align}
where in the laboratory frame $Q^2_{el}=|{\bf q}|^2-w_{el}^2$, $\omega_{el}=\sqrt{|{\bf q}|^2+m_A^2}-m_A$ is the energy
transfer corresponding to the elastic scattering, $m_A$ is the target nucleus mass and
\begin{align}
\epsilon_i=G_E^p(Q^2_{el})\frac{(1+\tau_{z_i})}{2}+G_E^n(Q^2_{el})\frac{(1-\tau_{z_i})}{2}\, .
\end{align}
The elastic form factor can be rewritten in terms of the Fourier transforms of the point-proton and nucleon
densities as
\begin{align}
F_L({\bf q})=&\frac{1}{Z}\frac{G_E^p(Q^2_{el})\tilde{\rho}_p(q)+G_E^n(Q^2_{el})\tilde{\rho}_n(q)}{\sqrt{1+Q^2_{el}/(4m^2)}} \, ,
\label{el:form:fact}
\end{align}
where 
\begin{align}
\tilde{\rho}_{p,n}(q)=&\int d^3r_1\ldots d^3r_A\, e^{i{\bf q}\cdot{\bf r}_1}\, \psi^\ast_0(r_1,\ldots,r_A)\nonumber\\
&\times\frac{1\pm\tau_{z_1}}{2}\psi_0(r_1,\ldots,r_A)\, \nonumber\\
=& \int d^3 r e^{i{\bf q}\cdot{\bf r}} \rho_{p,n}({\bf r})\, .
\label{FT:rho}
\end{align}
Note that the factor $\sqrt{1+Q^2_{el}/(4m^2)}$ in the denominator of Eq.~\eqref{el:form:fact} accounts for relativistic corrections 
to the charge operator.
Assuming that $\tilde{\rho}_{p}(q)=\tilde{\rho}_{n}(q)$, the charge distribution can be written as
\begin{align}
\rho_{ch}(r^\prime)=&\int\, \frac{d^3q}{(2\pi)^3} e^{-i{\bf q}\cdot {\bf r}^\prime} F_L({\bf q})\nonumber\\
=& \int\, \frac{d^3q}{(2\pi)^3} e^{-i{\bf q}\cdot {\bf r}^\prime} \frac{(G_E^p(Q^2_{el})+G_E^n(Q^2_{el}))\tilde{\rho}_p(q)}{\sqrt{1+Q^2_{el}/(4m^2)}}\, .
\label{rho:ch}
\end{align}
The probability of finding a nucleon in the nucleus with momentum ${\bf k}$ is proportional to its momentum distribution. 
The latter can be written in terms of the one-body density matrix of Eq.~\eqref{1b:dens} as
\begin{align}
n({\bf k})=\sum_{\alpha\beta}\tilde{\Phi}^\ast_\beta({\bf k})\tilde{\Phi}_\alpha({\bf k})\rho_{\alpha\beta}
\label{mom:dist}
\end{align}
where $\tilde\Phi_\alpha({\bf k})$ is the Fourier transform of the harmonic oscillator wave function
\begin{align}
\tilde{\Phi}_\alpha({\bf k})=\int d^3{r}\, e^{i {\bf k\, r}}\Phi_\alpha({\bf r})\, .
\label{HO:ft}
\end{align}
The momentum distribution is normalized as $\int d^3k\, n({\bf k})/(2\pi)^3=N$, with $N$ being the number of either protons $Z$, or neutrons $(A-Z)$.\\

The subtraction of the center of mass contribution from the wave function is a long standing issue affecting a number of nuclear many-body approaches relying on a single-nucleon basis expansions. Whilst for medium and heavy nuclei this correction can be safely neglected, the center of mass contribution strongly affects the results of light nuclei, such as $^4$He. 
In order to address this problem, we developed a Metropolis Monte Carlo (MMC) code, analogous to the one used in Variational Monte Carlo studies~\cite{Carlson:2014vla},  that allows us to single out
the center of mass contribution to the wave function in the calculation of the charge density and the momentum distribution. 
The wave function we used for the MMC is the Slater determinant obtained from the OpRS calculation, $|\psi_V\rangle=|\psi^{OpRS}_0\rangle$.
At variance with the fully correlated propagator of Eq.~\eqref{corr:prop}, the use of the unperturbed $g^{\rm{OpRS}}$,
(\textit{i.e.} a Slater determinant) allows for a unique definition of the wave function.
The spatial integrals of Eqs.~\eqref{FT:rho},~\eqref{mom:dist}, and~\eqref{HO:ft} have been performed using Metropolis Monte Carlo techniques \cite{Metropolis:1953am}. 
A sequence of points in the 3A-dimensional space denoted by ${\bf R}=\{{\bf r}_1,\ldots,{\bf r}_A\}$ are generated by sampling from the probability distribution 
\begin{align}
P({\bf R})=|\psi^{OpRS}_0({\bf R})|^2\ .
\end{align}
At each step of the calculation the center of mass contribution to the wave function is subtracted
computing the wave function and the expectation value in the intrinsic coordinates given by
\begin{align}
\tilde{\bf r}_i={\bf r}_i - {\bf R}_{cm}\ \ \ \ \ ,\ \ \ \ \ {\bf R}_{cm}=\frac{1}{A}\sum_i{{\bf r}_i}\, .
\end{align}
Hence, the identification of the intrinsic contribution is easily achieved  within Quantum Monte Carlo (QMC) algorithms~\cite{Schmidt:1999lik,Pieper:1990ji,Brida:2011yp,Pastore:2012rp}, since we always have access to the set of 3A-coordinates  of the constituent nucleons.

\section{The Impulse Approximation and convolution approach}
\label{IA:appr}

In the one-photon-exchange approximation, the  double differential electron-nucleus cross section takes the form
\begin{align}
\label{xsec}
\frac{d^2\sigma}{d E_{e^\prime} d\Omega_{e^\prime}}&=\frac{\alpha^2}{q^4}\frac{E_{e^\prime}}{E_e}L_{\mu\nu}W^{\mu\nu} \ ,
\end{align}
where $k_e=(E_e,{\bf k}_e)$ and $k_{e^\prime}=(E_{e^\prime},{\bf k}_{e^\prime})$ are the laboratory four-momenta of the incoming and outgoing electrons, respectively; $\alpha \simeq 1/137$ is the fine structure constant, $d\Omega_{e^\prime}$, the differential solid angle in the direction of ${\bf k}_{e^\prime}$, and $q=k_e - k_{e^\prime} =(\omega,{\bf q})$  the four momentum transfer.  
The leptonic tensor is given by
\begin{align}
  L^{\mu\nu}=2 \left( k_{e^\prime}^\mu k_e^\nu+ k_e^\mu k_{e^\prime}^\nu- g^{\mu\nu}k_{e^\prime}k_e \right)\,.  
\end{align}
The hadronic tensor is written in terms of matrix elements of the nuclear current operator between the target ground state and the hadronic final states as 
\begin{align}
\label{response:tensor}
W^{\mu \nu} =&\sum_f \langle  \psi_0^A| {J^\mu}^\dagger(q) | \psi_f^A \rangle  \langle  \psi_f^A| J^\nu(q) | \psi_0^A \rangle\nonumber\\
&\times \delta^{(4)}(P_0+q-P_f)   \ ,
\end{align}
where $|\psi_0^A \rangle$ and $|\psi_f^A \rangle$ denote the initial and final hadronic states with four-momenta $P_0 = ( E_0,{\bf p}_0 )$ and $P_f = (E_f,{\bf p}_f) $, while $J(q)$ is the electromagnetic nuclear current operator.

At relatively large momentum transfer, $|{\bf q}| \gtrsim 500$ MeV, the Impulse Approximation (IA) can be safely applied. Within this approximation the interaction
between the struck nucleon and the spectator (A-1) particles is neglected~\cite{Benhar:2006wy, Benhar:2015wva}. The nuclear current operator reduces to a sum of
one-body terms, $J(q)=\sum_i j_i(q)$ and the hadronic final state factorizes as
\begin{align}  
|f\rangle \rightarrow |p\rangle \otimes |\psi_f^{A-1}\rangle\, .
\end{align}
In the above equation $|p\rangle$ denotes the final-state nucleon with momentum ${\bf p}$, while $|\psi_f^{A-1}\rangle$ describes the \hbox{$(A-1)$-body} spectator system. 
Its energy and recoiling momentum are fixed by energy and momentum conservation, yielding
\begin{align}
E_f^{A-1}=&\omega +E^A_0-e({\bf p})\, ,\ \ \ \ {\bf P}^{A-1}_f={\bf q}-{\bf p}\, .
\end{align}
Using the factorization ansatz and inserting a single-nucleon completeness relation, the matrix element of the current operator can be written as
\begin{align}
\langle \psi_0^A|J^\mu|\psi_f^A\rangle \rightarrow \sum_k \langle \psi_0^A| [|k\rangle \otimes |\psi_f^{A-1}\rangle]\langle k|\sum_i j^\mu_i|p\rangle\, .
\end{align}
Substituting the last equation in Eq.~\eqref{response:tensor}, the contribution to the hadron tensor is given by
\begin{align}
W^{\mu\nu}({\bf q},\omega)&=\sum_{p,k}\sum_{f,i}\, \langle k | {j_{i}^\mu}^\dagger |p \rangle \langle p |  j_{i}^\nu | k \rangle\nonumber\\
&\times \langle \psi_0^A| [ |\psi_f^{A-1}\rangle \otimes |k \rangle ]  [ \,\langle \psi_f^{A-1} | \otimes \langle k| ]  |  \psi_0^A\rangle \nonumber\\
&\times \delta(\omega-e(\mathbf{p}) - E_{f}^{A-1} +E^A_0)\, .
\end{align}
Momentum conservation in the single-nucleon vertex implies ${\bf p}={\bf k}+{\bf q}$. The one-body current operator 
can be written as
\begin{align}
j^{\mu}_i=& \Big[ F_{1i}\gamma^\mu+i \frac{F_{2i}}{2m}\sigma^{\mu\nu}q_\nu\Big]
\end{align}
where 
\begin{align}
F_{1,2i}=&\frac{(F^S_{1,2}+F^V_{1,2}\tau_{iz})}{2}\, ,
\end{align}
and $F^{S(V)}=F^{p}\pm F^{n}$. The latter are defined in terms of the electric and magnetic form factors via
\begin{align}
F^S_{1}=&\frac{G^S_{E}+\tau G^S_{M}}{1+\tau}\, ,\nonumber\\
F^S_{2}=&\frac{G^S_{M}-G^S_{E}}{1+\tau}\, ,
\end{align}
where $\tau=-q^2/(4m^2)$.
Finally, using the identity
\begin{align}
\delta(\omega&-e({\bf p})-E^{A-1}_f+E^A_0)=\nonumber\\
&   \int dE\, \delta(\omega+E-e({\bf p})) \, \delta(E+E^{A-1}_f-E^A_0)  \; ,
\end{align}
we can rewrite the hadron tensor as
\begin{align}
W^{\mu\nu}({\bf q},\omega)=& \int \frac{d^3k}{(2\pi)^3} dE P_h({\bf k},E)\frac{m^2}{e({\bf k})e({\bf k+q})}\nonumber\\
&\times  \sum_{i}\, \langle k | {j_{i}^\mu}^\dagger |k+q \rangle \langle k+q |  j_{i}^\nu | k \rangle\nonumber\\
& \times \delta(\omega+E-e(\mathbf{k+q}))\, ,
\label{had:tens2}
\end{align}
where the factors  $m/e({\bf k})$ and $m/e({\bf k+q})$ have to be included to account for the implicit covariant 
normalization of the four-spinors of the initial and final nucleons in the matrix elements of the relativistic current.\\
The hole spectral function
\begin{align}
P_h(\mathbf{k},E)&=\sum_f |\langle \psi_0^A| [|\mathbf{k}\rangle \otimes |\psi_f^{A-1}\rangle]|^2\nonumber\\
&\times\delta(E+E_{f}^{A-1}-E^A_0)\, .
\end{align}
gives the probability distribution of removing a nucleon with momentum ${\bf k}$ from the target nucleus, 
leaving the residual $(A-1)$ system with an excitation energy $E$.
Note that in Eq.~\eqref{had:tens2} we neglected Coulomb interactions and the other (small) isospin-breaking terms and made the assumption, largely justified in the case of closed shell nuclei, 
that the proton and neutron spectral functions are identical.\\
Rewriting the nuclear matrix element as
\begin{align}
[ \,\langle \psi_f^{A-1} | \otimes \langle k| ]  |  \psi_0^A\rangle &= \sum_\alpha {\cal Y}^k_\alpha \tilde{\Phi}_\alpha({\bf k}) \nonumber\\
&= \sum_\alpha \tilde{\Phi}_\alpha({\bf k}) \langle \psi_f^{A-1} | a_\alpha |\psi_0^A\rangle\, ,
\end{align}
we recover the more familiar expression of the spectral function written as the imaginary part of the Green's function describing the propagation of a hole state
\begin{align}
P_h(\mathbf{k},E)&=\frac{1}{\pi}\sum_{\alpha\beta}\tilde{\Phi}^\ast_\beta({\bf k})\tilde{\Phi}_\alpha({\bf k})\nonumber\\
&\times\text{Im}\langle \psi_0^A| a_\beta^\dagger  \frac{1}{E+(H-E^A_0)-i\epsilon} a_\alpha|\psi_0^A\rangle\, .
\label{eq:Ph}
\end{align}

In the kinematical region in which the interactions between the struck particle and the spectator system can not be neglected, the IA results have to be modified to include the effect of FSI. Following Refs.~\cite{Ankowski:2014yfa,Benhar:1991af}, the multiple scatterings that the struck particle undergoes during its propagation through the nuclear medium are taken into account through a convolution scheme. The IA responses are folded with a function $f_\mathbf{k+q}$, normalized as 
\begin{equation}
\int_{-\infty}^{+\infty} d\omega f_{\bf k+q}(\omega) = 1\ .
\label{fold:func}
\end{equation}

The double differential cross section is then given by
\begin{align}
\Big(\frac{d^2\sigma }{ d E_{e^\prime} d\Omega_{e^\prime}}&\Big)_{FSI}= \int \frac{d^3k}{(2\pi)^3}\, dE \int d\omega^\prime\,f_{\bf k+q}(\omega-\omega^\prime)\nonumber\\
& \times  \frac{m}{e({\bf k})}\frac{m}{e({\bf k+q})}P_{h}({\bf k},E)\frac{\alpha^2}{q^4}\frac{E_{e^\prime}}{E_e}\nonumber\\
&\times L_{\mu\nu}\sum_{i}\, \langle k | \left(j_{i}^\mu\right)^\dagger |k+q \rangle \langle k+q |  j_{i}^\nu | k \rangle\nonumber\\
&\times \delta(\omega^\prime +E-\tilde{e}({\bf k+q})) \theta(|{\bf k+q}|-p_F)\, .
\label{dens:resp:folding}
\end{align}

In the last equation we modified the energy spectrum of the struck nucleon 
\begin{equation}
\tilde{e}({\bf k+q})=e({\bf k+q})+ U\left(t_{\rm kin}({\bf k+q})\right)\, 
\end{equation}
by considering the real part of the optical potential $U$ derived from the Dirac phenomenological fit of Ref.~\cite{Cooper:1993nx}. This allows to describe the propagation of the knocked-out particle in the mean-field generated by the spectator system.

\section{results}
\label{results}
\begin{figure}[]
\centering
\includegraphics[scale=0.675]{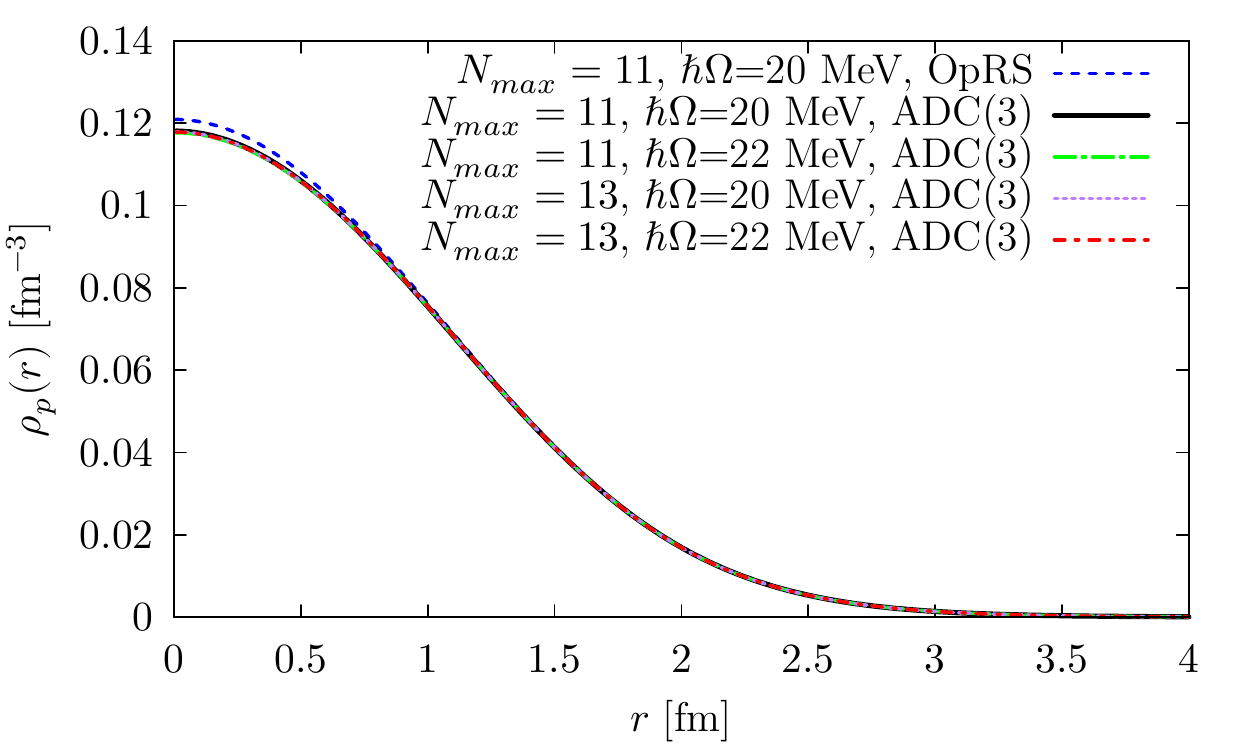}
\caption{Point proton densities in $^4$He. The dashed (blue) line corresponds to the OpRS derived for $\nmax=11$ and
$\hbar\Omega=20$ MeV. The other lines have been obtained using the SCGF full propagator for $\nmax=$11, 13 and $\hbar\Omega=$20, 22~MeV.}
\label{rho:p:He4}
\end{figure}

Our calculations have been performed using the NNLO$_{\rm sat}$ chiral interaction~\cite{Ekstrom:2015rta}, which 
was specifically designed to accurately describe both binding energies and nuclear radii of mid-mass nuclei~\cite{Ruiz2015Nature,Lapoux:2016exf}.
In Fig.~\ref{rho:p:He4} we analyze the convergence of the SCGF-ADC(3) point-proton densities of $^4$He
with respect to the oscillator frequency ($\hbar\Omega$) and the size of the model space ($\nmax$). The different lines almost superimpose, indicating that for $\hbar\Omega\approx$~20 MeV and 
$\nmax\geq$11 the calculation converges and no longer depends on the oscillator parameters.
The density calculated from the OpRS is also displayed. The nice agreement with the SCGF-ADC(3) curves follows from the requirement that the single particle energies and overlap functions in the OpRS propagator are chosen to approximate at best the true (correlated) one-body density.

\begin{figure}[]
\centering
\includegraphics[scale=0.675]{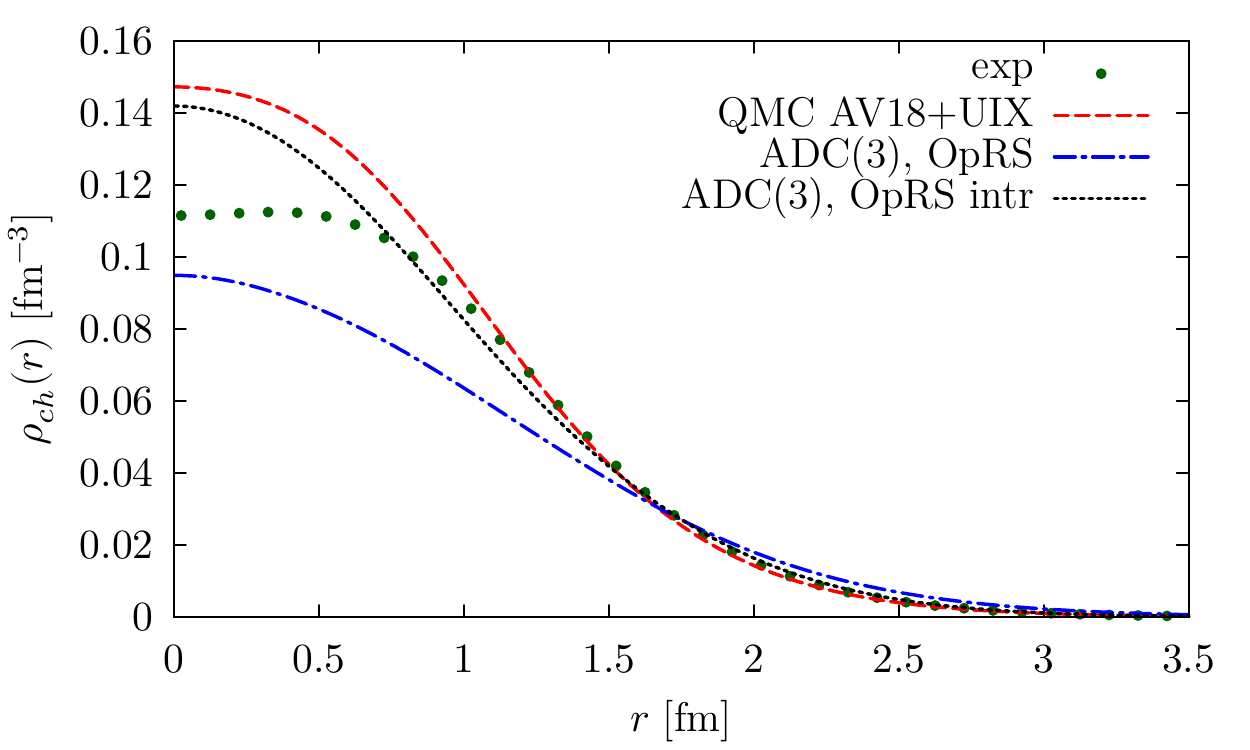}
\caption{Charge densities of $^4$He. The (green) dots have been obtained using the ``Sum-of-Gaussians'' parametrization of the charge densities given in Ref.~\cite{DeJager:1987qc}.
The dashed (red) line refers to the QMC calculation of Ref.~\cite{Lonardoni:2017egu} that used the AV18+UIX two- and three-body interactions.
The dot-dashed (blue) line corresponds to the same OpRS propagator shown in Fig.~\ref{rho:p:He4}, while in the short-dashed (black) line the center-of-mass contamination has been subtracted from the OpRS wave function by means a MMC calculation.}
\label{rho:ch:He4}
\end{figure}

The charge densities in $^4$He can be obtained from the point-proton densities through Eqs.~\eqref{FT:rho} and~\eqref{rho:ch}. In Fig.~\ref{rho:ch:He4}
we compare the experimental charge density determined through the ``Sum-of-Gaussians'' parametrization given in Ref.~\cite{DeJager:1987qc} with those obtained from the QMC results of Ref.~\cite{Lonardoni:2017egu} and from the OpRS calculated in the present work.
For the latter, we display both the result already shown in Fig.~\ref{rho:p:He4} and the distribution obtained after subtracting the center of mass effect with the MMC algorithm outlined in Sec.~\ref{sec:SCGF}.
When the center of mass contamination is subtracted, we obtain the short-dashed (black) line. 
The comparison with the total OpRS results, corresponding to the dot-dashed (blue) line, clearly shows that for $^4$He the center of mass contribution is sizeable and can not be neglected. 
The use of the intrinsic wave function yields a strong enhancement of the charge density, which turns out to be very close to the QMC result. 
Note that the discrepancy between the experiment and the intrinsic OpRS and QMC calculations is motivated by the absence of the 
 two-body meson exchange current contributions.
These are known to have little effect on larger nuclei such as $^{16}$O but their inclusion is fundamental in order to correctly reproduce the $^4$He elastic form factor, from which the charge densities are extracted ~\cite{Lonardoni:2017egu,Carlson:2014vla,Marcucci:2015rca,Lonardoni:2018nob}.

\begin{figure}[]
\centering
\includegraphics[scale=0.675]{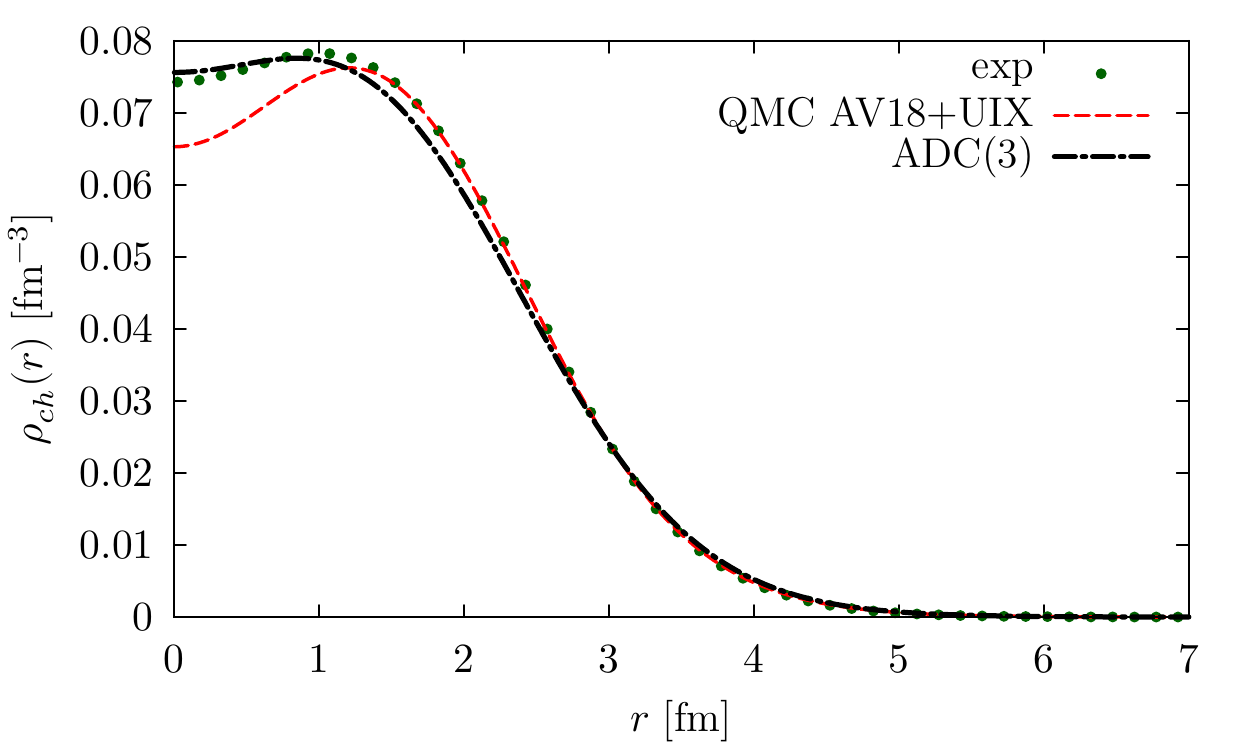}
\caption{Charge densities in $^{16}$O. The (green) dots and the dashed (red) line are the same as Fig.~\ref{rho:ch:He4}. The dot-dashed (black) line 
corresponds to the full SCGF density calculated at the ADC(3) level.}
\label{rho:ch:O16}
\end{figure}

For medium-mass nuclei, the center of mass corrections are known to be less significant. Therefore, in Fig.~\ref{rho:ch:O16} we compare the experimental charge density in $^{16}$O with the full SCGF-ADC(3) 
and the QMC calculations. There is an overall nice agreement between the theoretical curves. The SCGF-ADC(3) results perfectly reproduce the experimental points, confirming the goodness of
the $\rm{NNLO_{sat}}$ potential which was fitted to reproduce the experimental radius of $^{16}$O.

\begin{figure}[]
\centering
\includegraphics[scale=0.675]{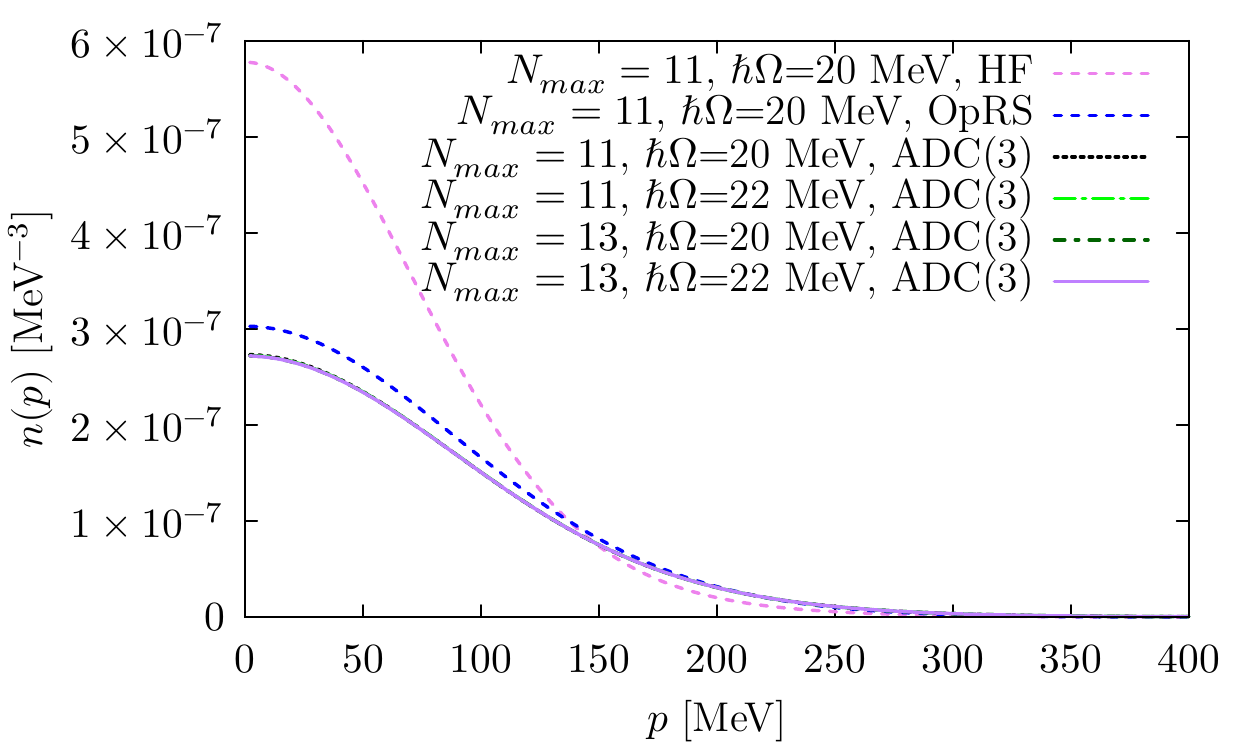}
\caption{Momentum distributions in $^4$He. The dashed (purple and blue) lines corresponds to the HF and OpRS  approximations for $\nmax=11$ and
$\hbar\Omega=20$ MeV. The other lines have been obtained using the SCGF full propagator at the ADC(3) level for $\nmax=$11, 13 and $\hbar\Omega=$20, 22 MeV.}
\label{nk:compare:Nw}
\end{figure}

\begin{figure}[]
\centering
\includegraphics[scale=0.675]{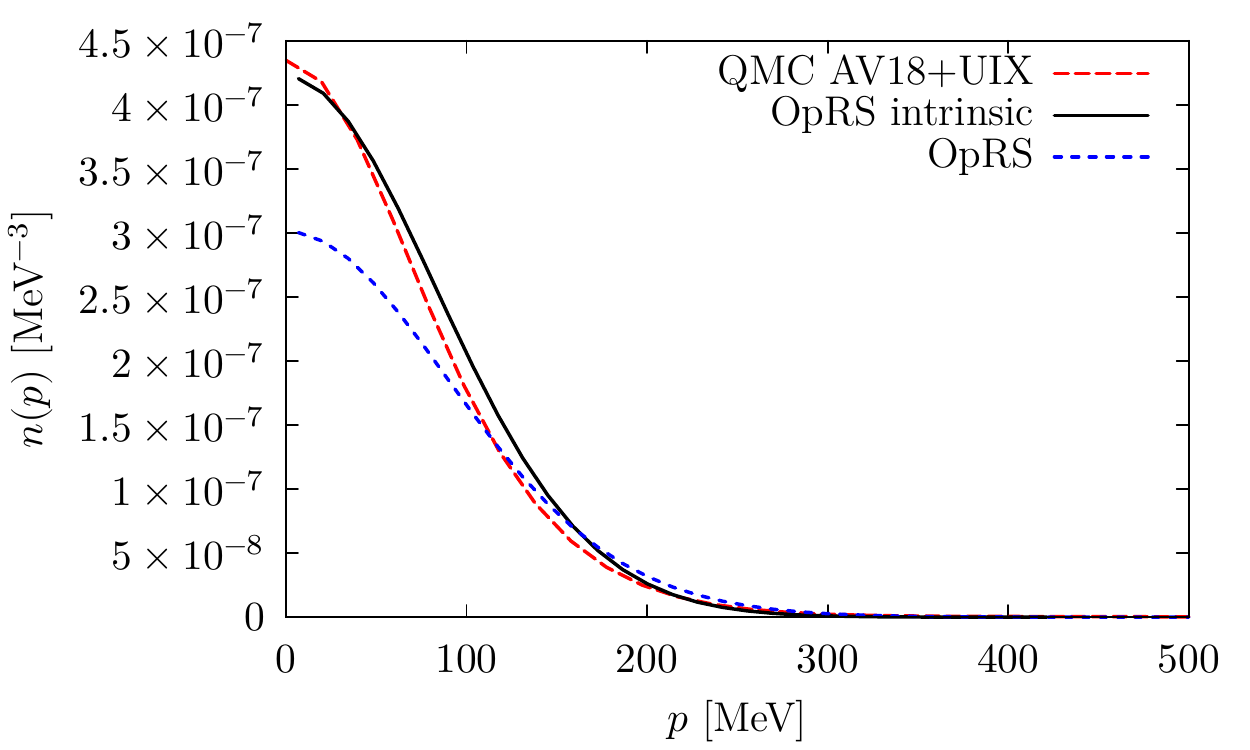}
\caption{
Momentum distributions of $^4$He. The dashed (red) line corresponds to the QMC calculation~\cite{Lonardoni:2017egu} while the short-dashed (blue) and solid (black) lines correspond to the
total and intrinsic OpRS results, respectively.
}
\label{nk:compare:He4}
\end{figure}
\begin{figure}[]
\centering
\includegraphics[scale=0.675]{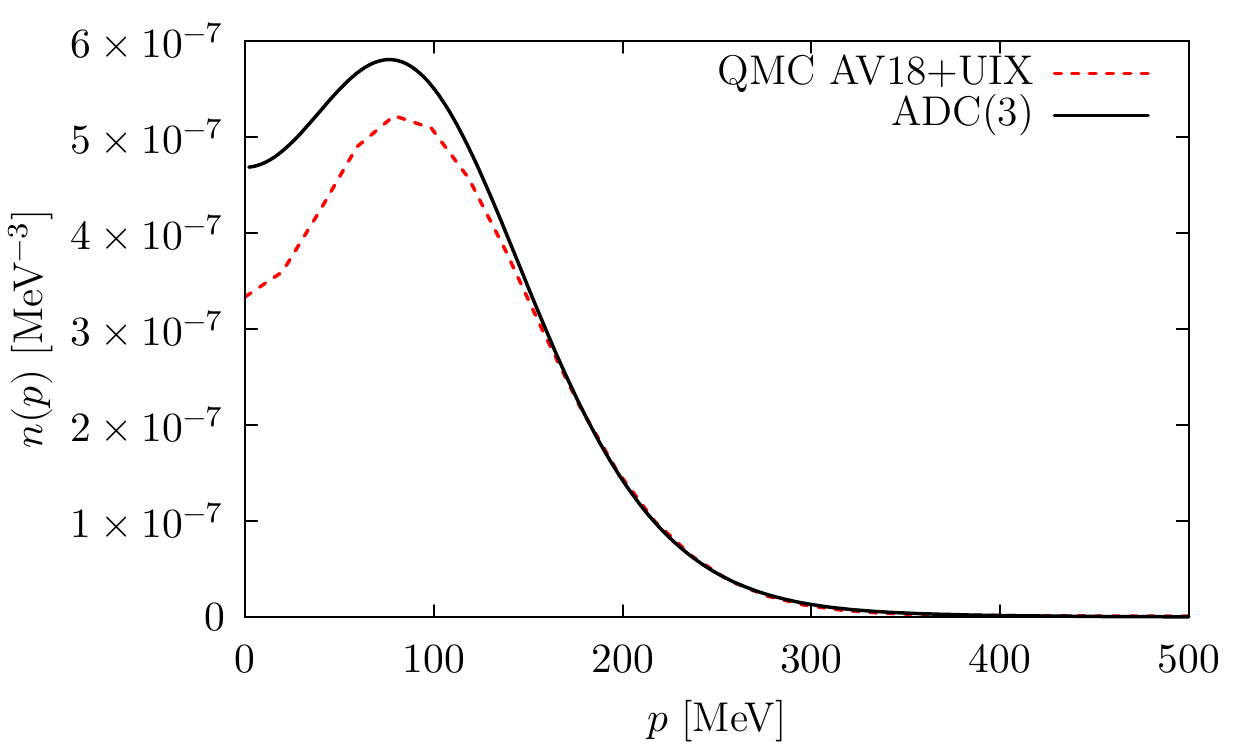}
\includegraphics[scale=0.675]{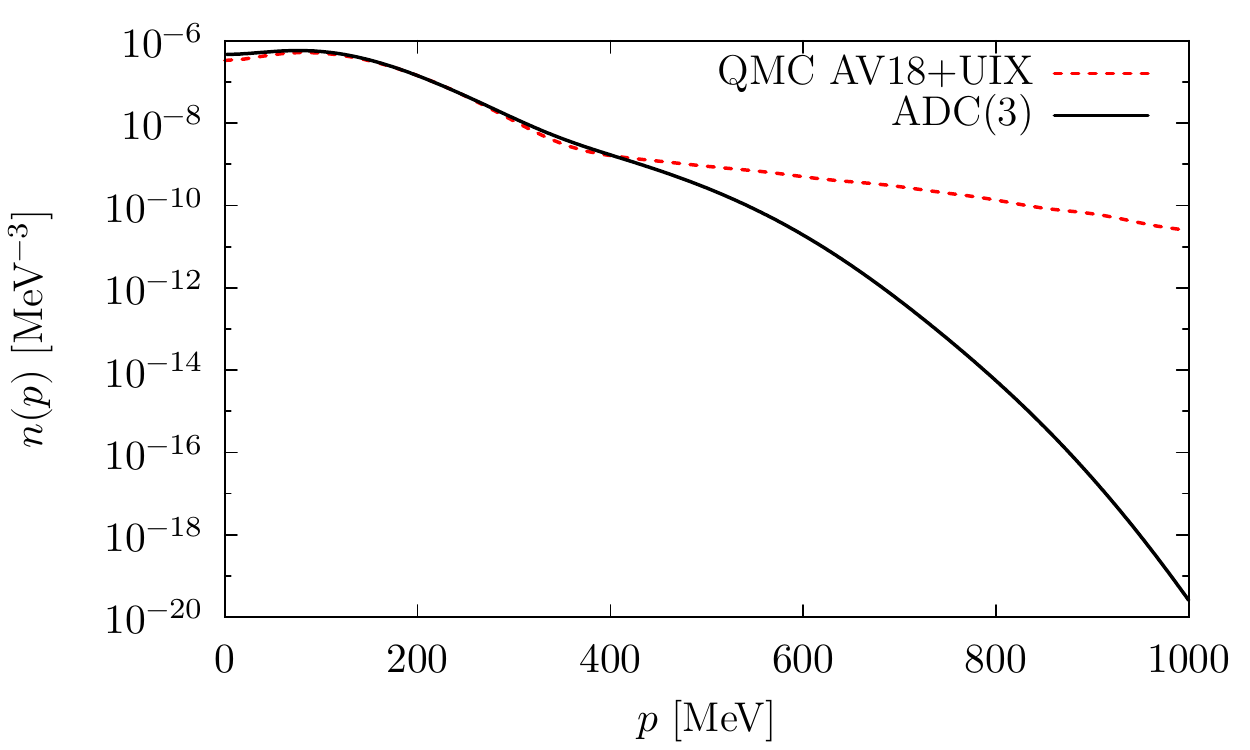}
\caption{Computed momentum distributions of $^{16}$O. The dashed (red) and solid (black) lines are obtained within QMC~\cite{Lonardoni:2017egu} and SCGF-ADC(3) approaches, respectively. In the lower panel, a logarithmic scale has been used to demonstrate the weak tail at large momenta that arises from the soft chiral interaction adopted in the SCGF-ADC(3) calculation. }
\label{nk:compare:O16:log}
\end{figure}

\begin{figure*}[]
\includegraphics[scale=0.575]{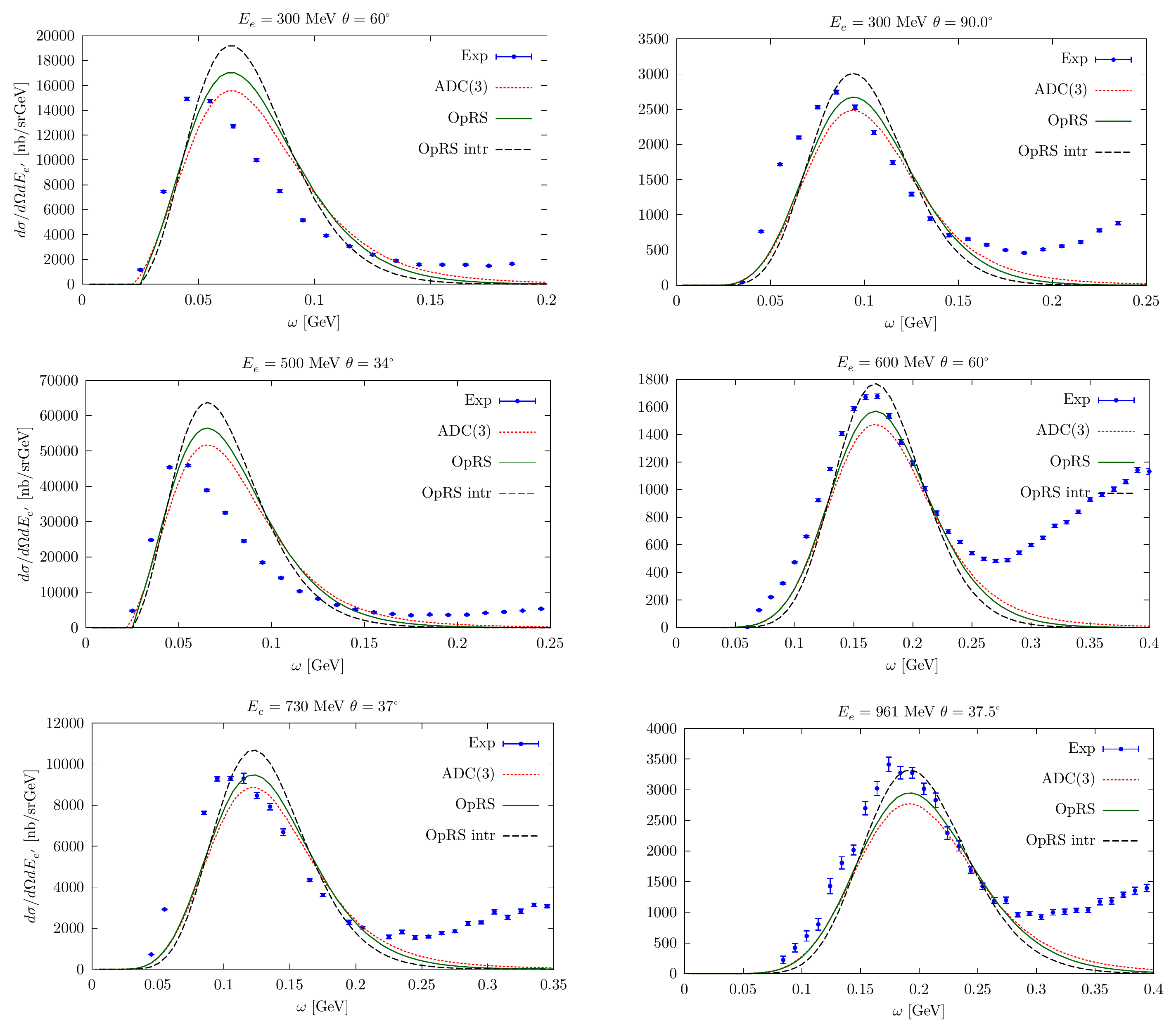}
\caption{Double-differential electron-$^4$He cross sections for different values of incident electron energy 
and scattering angle. The dotted (red) curve have been obtained using the SCGF-ADC(3) propagator while the solid (green) and dashed (black) line corresponds to the total and the intrinsic
OpRS results, respectively. The experimental data are taken from Ref.~\cite{Benhar:2006er}. 
}
\label{cross:sec:4He}
\end{figure*}

\begin{figure*}[]
\includegraphics[scale=0.575]{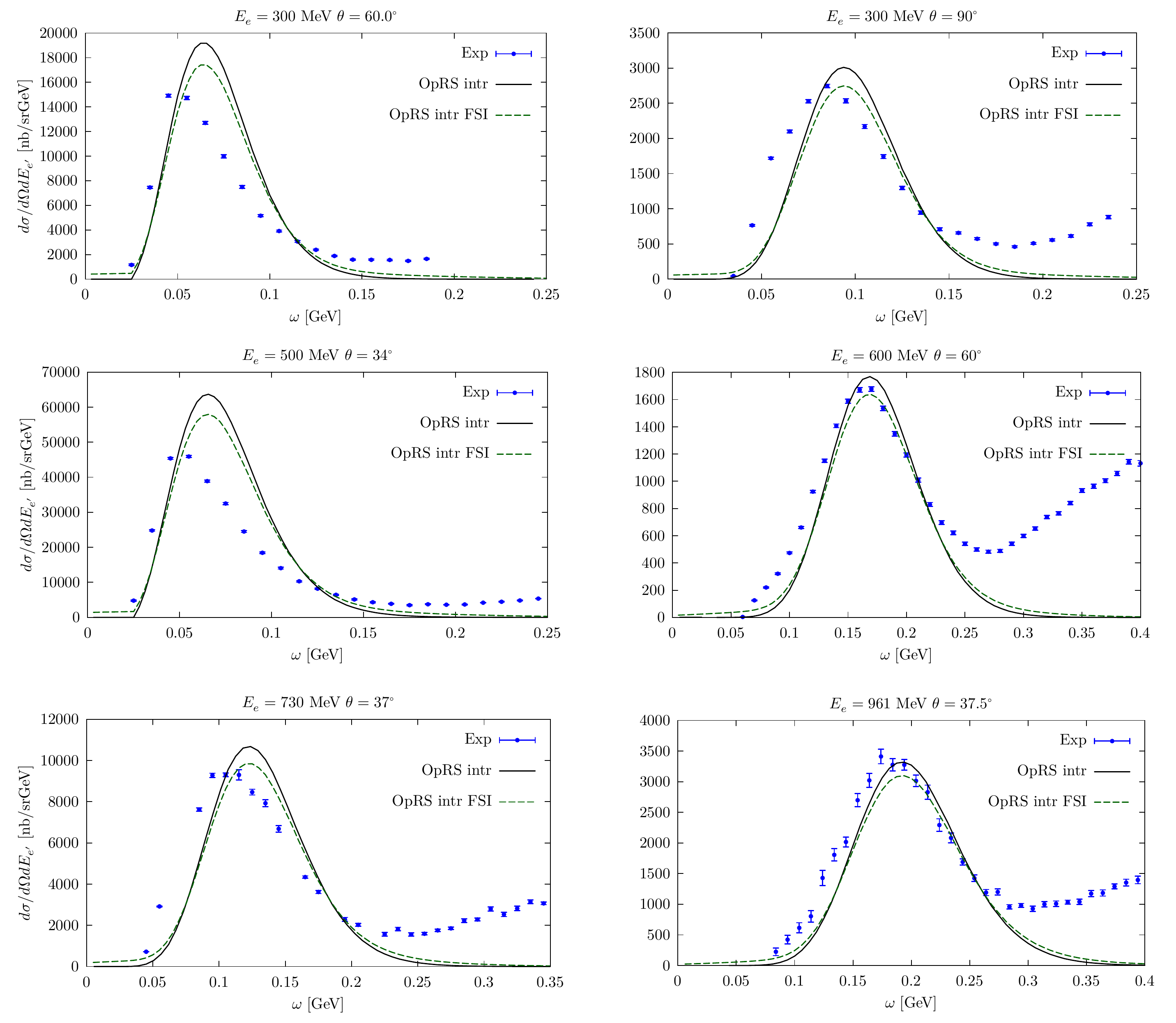}
\caption{Same double-differential cross sections as in Fig.~\ref{cross:sec:4He} but only for the intrinsic OpRS calculation. The solid (black) line corresponds to neglecting FSI, while  the dashed (green) one has been obtained including FSI corrections. 
 }
\label{cross:sec:4He:fsi}
\end{figure*}

\begin{figure*}[]
\includegraphics[scale=0.575]{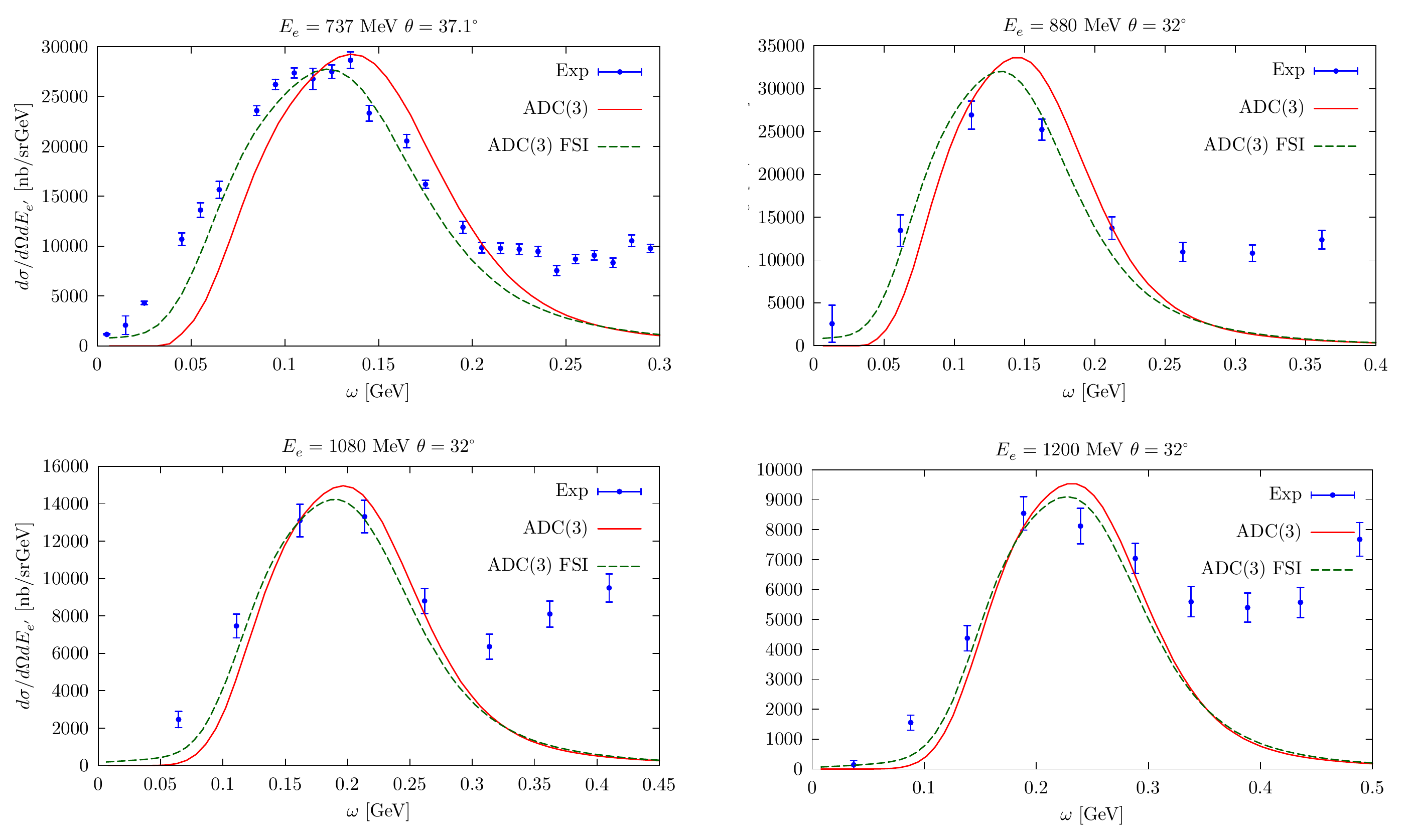}
\caption{Double-differential electron-$^{16}$O cross sections for different values of incident electron energy 
and scattering angle. The solid (red) line corresponds to the SCGF-ADC(3) results and the dashed (green) one has been obtained including FSI corrections. The experimental data are taken from Ref.~\cite{Benhar:2006er}. 
}
\label{cross:sec:16O}
\end{figure*}

The different ADC(3) curves in Fig.~\ref{nk:compare:Nw} correspond to the single-nucleon momentum distribution 
of $^4$He obtained by varying the oscillator frequency and the size of the model space.
As it has been already observed in Fig.~\ref{rho:p:He4}, for these values of $\hbar\Omega$ and $\nmax$ the calculation 
is independent of the oscillator parameters.
The OpRS result, corresponding to the dashed (blue) line, correctly follows that of the dressed ADC(3) propagator, 
although the agreement is not as close as in Fig.~\ref{rho:p:He4}.
The distribution obtained from the Hartree-Fock state are much different and they are also displayed to further emphasize the importance of nucleon-nucleon correlations. It is 
clearly visible that their correct inclusion leads to a strong reduction of
the momentum distribution in the region of low momentum and to the appearance of
a tail for values of $p$ well above the Fermi momentum.

In Fig.~\ref{nk:compare:He4} we benchmark the intrinsic and uncorrected OpRS single-nucleon momentum distribution of $^4$He with the QMC calculation of Ref.~\cite{Lonardoni:2017egu}. 
Note that, also in this case the subtraction of the center of mass component has a sizable effect, which is crucial for recovering the agreement with the intrinsic QMC results. 

The $^{16}$O single-nucleon momentum distributions  obtained within the SCGF-ADC(3) and QMC approach are compared in Fig.~\ref{nk:compare:O16:log}. 
In the upper panel the dashed (red) line, corresponding to the QMC calculation, is
found to be below the SCGF-ADC(3) results for low momenta. This is likely to be ascribed to the different choice made for the potentials. In fact, the NNLO$_{\rm{sat}}$ is much softer than
the AV18+UIX potential adopted in the QMC study. 
The use of an hard potential implies the presence of stronger high momentum
components in the nuclear wave function. This becomes evident in the lower panel where the logarithmic scale has been used to show the differences in the tails.
While the QMC momentum distribution exhibits a long tail extending to $p>1$ GeV, the use of an harmonic oscillator model space and the softer potential adopted in our calculations strongly reduce the SCGF-ADC(3) momentum distribution in the high momentum region.

Fig.~\ref{cross:sec:4He} shows the electron-$^4$He inclusive double-differential cross sections at different values of $E_e$ and $\theta_e$. The curves are obtained from the full SCGF-ADC(3) spectral function,  from its OpRS approximation and from the intrinsic OpRS.
The SCGF-ADC(3) cross-section represented by the dashed (red) line is quenched with respect to the solid (green) line that refers to the uncorrected OpRS.
This has to be attributed to the different behavior of the curves displayed in Fig.~\ref{nk:compare:Nw}. Whilst the OpRS wave functions are built  to reproduce lowest energy momenta of the ADC(3) propagator---which optimises the quasiparticle energies and strength near the Fermi surface---this leaves small discrepancies in the single-nucleon momentum distribution.
The comparison between the solid (green) and dashed (black) curve clearly shows that the subtraction of the center of mass component from the wave function
leads to a reduction of the width and an enhancement of the quasielastic peak. Since this strongly affects  
the cross section in all the kinematical setups that we considered, we applied FSI corrections only to the intrinsic OpRS calculation. 
In order to do it, we follow the approach outlined in Sec.~\ref{IA:appr}, with the difference that
 the optical potential has been disregarded in the energy conserving $\delta$-function since to the best of our knowledge neither the $^3$H-p nor the $^3$He-n optical potentials are present in the literature. The results are shown in Fig.~\ref{cross:sec:4He:fsi}.
The convolution of the OpRS cross section with the folding function of Eq.~\eqref{fold:func} leads to a redistribution of the strength, which quenches the peak and enhances the tails.
For $E_e= 300$ MeV, $\theta=60^{\circ}$, and $E_e= 500$ MeV, $\theta=34^{\circ}$ the OpRS intrinsic calculation overestimates the 
data. Moreover, in all the kinematical configurations under consideration the position of the quasielastic peak is not correctly reproduced. This is
likely to be ascribed to the approximate procedure we adopted to account for FSI effects, \textit{i.e.} we neglected the real part of the optical
potential.  Its inclusion would shift the cross section towards lower values of $\omega$ possibly
 improving the agreement with the experimental data. 

In Fig.~\ref{cross:sec:16O} we compare the experimental data of the inclusive double-differential electron-$^{16}$O cross sections 
as computed from the fully correlated SCGF-ADC(3) spectral function.
In the dashed (green) curve FSI effects have
been  implemented in full, yielding a very nice agreement with the data. 
In particular, the inclusion of the real part of the optical potential in the final state nucleon energy
shifts the cross sections towards lower values of $\omega$ and the
quasielastic-peak position is correctly reproduced.

\section{Conclusions}
\label{concl}
We used the \textit{ab initio} SCGF approach to compute the single-particle propagators of closed shell $^4$He and $^{16}$O nuclei. The calculations were based on the the NNLO$_{\rm {sat}}$ chiral interaction since this is cable to describe simultaneously binding energies and nuclear radii of medium-mass nuclei.
We have gauged the residual center of mass contribution to the $^4$He wave function by 
developing a MMC  algorithm which exploits the OpRS single-particle propagator.
The resulting intrinsic charge density in $^4$He has been computed and compared
with both the QMC calculation of Ref.~\cite{Lonardoni:2017egu} and the experimental data. 
The subtraction of the center of mass contribution turns out to be crucial in order to obtain  correct predictions for this light nucleus. 
The same pattern has also been observed in the single-nucleon momentum distribution; whilst the total OpRS sizably underestimates the QMC calculation, 
a very nice agreement is found between the QMC and the results from the intrinsic OpRS. \\
In the analysis of the charge density in $^{16}$O, the full SCGF-ADC(3) calculation has been compared with the
experimental curve.
Since the radius of this nucleus has been used to fit the NNLO$_{\rm {sat}}$  potential, 
the nearly perfect agreement with the empirical charge density is not surprising. However, the very good comparison with 
experimental cross sections corroborates the choice of the interaction for future studies of lepton-nucleus scattering. 
The origin of the discrepancies between the single-nucleon momentum distributions obtained from SCGF-ADC(3) and QMC approaches
has to be attributed to the softness of the NNLO$_{\rm {sat}}$ interaction.
Although the two approaches provide very similar results in the region of low- and moderate-momentum, the use of 
an hard potential, such as AV18+UIX, implies a stronger nuclear interaction between large momenta. 
This manifests itself into the appearance of very high-momentum
tails in the momentum distribution which are not as pronounced for the NNLO$_{\rm {sat}}$ chiral force.

We employed the IA approach to perform the calculation of inclusive electromagnetic cross sections
which exploits SCGF spectral functions.
The electron-$^4$He double differential cross section corresponding to the intrinsic OpRS wave function
sizably differs from the OpRS in which the contamination of the center of mass is still present.
This indicates that the spurious effect of the center of mass can not be neglected in light nuclei.
For this reason, we restricted the discussion of FSI effects to the sole
OpRS intrinsic calculations. We observed that the convolution with the folding function of Ref.~\cite{Ankowski:2014yfa,Benhar:1991af}
yields a redistribution of the strength of the cross section. However, disregarding the real part of the optical
potential in the energy conserving delta function prevents a good agreement with the 
data for the different kinematical setup analyzed. 

Fully satisfactory results have been obtained for the electron-$^{16}$O double differential cross section, where the IA 
calculation has been supplemented by FSI. 
Our findings indicate that the SCGF approach provides accurate predictions for 
medium-mass nuclei and their interaction with an electron probe.
The extension to the electroweak sector will be the subject of a future work. 
Moreover, exploiting the Gorkov formalism we will be able to provide
 valuable results for open shell nuclei~\cite{Soma:2013xha} which will be crucial 
 in the data analysis of future neutrino experiments, such as DUNE.

\section{Acknowledgements} 
We are deeply indebted to O. Benhar, D. Lonardoni, and A. Lovato for many illuminating discussions and to P.~Navr\'{a}til for providing the matrix elements of NNLO$_{\rm {sat}}$ interaction.
This research has been supported by the Centro 
Nazionale delle Ricerche (CNR) and the Royal Society under the CNR-Royal Society International Fellowship scheme NF161046 and by the United Kingdom Science and Technology Facilities Council (STFC) under Grants No.~ST/L005743/1 and ST/L005816/1. 
Calculations were performed at the DiRAC Complexity system at the University of Leicester (BIS National \hbox{E-infrastructure} capital Grant No. ST/K000373/1 and STFC Grant No. ST/K0003259/1).

\bibliography{biblio}

\begin{thebibliography}{45}%
\makeatletter
\providecommand \@ifxundefined [1]{%
 \@ifx{#1\undefined}
}%
\providecommand \@ifnum [1]{%
 \ifnum #1\expandafter \@firstoftwo
 \else \expandafter \@secondoftwo
 \fi
}%
\providecommand \@ifx [1]{%
 \ifx #1\expandafter \@firstoftwo
 \else \expandafter \@secondoftwo
 \fi
}%
\providecommand \natexlab [1]{#1}%
\providecommand \enquote  [1]{``#1''}%
\providecommand \bibnamefont  [1]{#1}%
\providecommand \bibfnamefont [1]{#1}%
\providecommand \citenamefont [1]{#1}%
\providecommand \href@noop [0]{\@secondoftwo}%
\providecommand \href [0]{\begingroup \@sanitize@url \@href}%
\providecommand \@href[1]{\@@startlink{#1}\@@href}%
\providecommand \@@href[1]{\endgroup#1\@@endlink}%
\providecommand \@sanitize@url [0]{\catcode `\\12\catcode `\$12\catcode
  `\&12\catcode `\#12\catcode `\^12\catcode `\_12\catcode `\%12\relax}%
\providecommand \@@startlink[1]{}%
\providecommand \@@endlink[0]{}%
\providecommand \url  [0]{\begingroup\@sanitize@url \@url }%
\providecommand \@url [1]{\endgroup\@href {#1}{\urlprefix }}%
\providecommand \urlprefix  [0]{URL }%
\providecommand \Eprint [0]{\href }%
\providecommand \doibase [0]{http://dx.doi.org/}%
\providecommand \selectlanguage [0]{\@gobble}%
\providecommand \bibinfo  [0]{\@secondoftwo}%
\providecommand \bibfield  [0]{\@secondoftwo}%
\providecommand \translation [1]{[#1]}%
\providecommand \BibitemOpen [0]{}%
\providecommand \bibitemStop [0]{}%
\providecommand \bibitemNoStop [0]{.\EOS\space}%
\providecommand \EOS [0]{\spacefactor3000\relax}%
\providecommand \BibitemShut  [1]{\csname bibitem#1\endcsname}%
\let\auto@bib@innerbib\@empty
\bibitem [{\citenamefont {Gran}\ \emph {et~al.}(2006)\citenamefont {Gran} \emph
  {et~al.}}]{Gran:2006jn}%
  \BibitemOpen
  \bibfield  {author} {\bibinfo {author} {\bibfnamefont {R.}~\bibnamefont
  {Gran}} \emph {et~al.} (\bibinfo {collaboration} {K2K}),\ }\href {\doibase
  10.1103/PhysRevD.74.052002} {\bibfield  {journal} {\bibinfo  {journal} {Phys.
  Rev.}\ }\textbf {\bibinfo {volume} {D74}},\ \bibinfo {pages} {052002}
  (\bibinfo {year} {2006})},\ \Eprint {http://arxiv.org/abs/hep-ex/0603034}
  {arXiv:hep-ex/0603034 [hep-ex]} \BibitemShut {NoStop}%
\bibitem [{\citenamefont {Aguilar-Arevalo}\ \emph {et~al.}(2008)\citenamefont
  {Aguilar-Arevalo} \emph {et~al.}}]{AguilarArevalo:2007ab}%
  \BibitemOpen
  \bibfield  {author} {\bibinfo {author} {\bibfnamefont {A.~A.}\ \bibnamefont
  {Aguilar-Arevalo}} \emph {et~al.} (\bibinfo {collaboration} {MiniBooNE}),\
  }\href {\doibase 10.1103/PhysRevLett.100.032301} {\bibfield  {journal}
  {\bibinfo  {journal} {Phys. Rev. Lett.}\ }\textbf {\bibinfo {volume} {100}},\
  \bibinfo {pages} {032301} (\bibinfo {year} {2008})},\ \Eprint
  {http://arxiv.org/abs/0706.0926} {arXiv:0706.0926 [hep-ex]} \BibitemShut
  {NoStop}%
\bibitem [{\citenamefont {Lyubushkin}\ \emph {et~al.}(2009)\citenamefont
  {Lyubushkin} \emph {et~al.}}]{Lyubushkin:2008pe}%
  \BibitemOpen
  \bibfield  {author} {\bibinfo {author} {\bibfnamefont {V.}~\bibnamefont
  {Lyubushkin}} \emph {et~al.} (\bibinfo {collaboration} {NOMAD}),\ }\href
  {\doibase 10.1140/epjc/s10052-009-1113-0} {\bibfield  {journal} {\bibinfo
  {journal} {Eur. Phys. J.}\ }\textbf {\bibinfo {volume} {C63}},\ \bibinfo
  {pages} {355} (\bibinfo {year} {2009})},\ \Eprint
  {http://arxiv.org/abs/0812.4543} {arXiv:0812.4543 [hep-ex]} \BibitemShut
  {NoStop}%
\bibitem [{\citenamefont {Dickhoff}\ and\ \citenamefont
  {Barbieri}(2004)}]{Dickhoff:2004xx}%
  \BibitemOpen
  \bibfield  {author} {\bibinfo {author} {\bibfnamefont {W.~H.}\ \bibnamefont
  {Dickhoff}}\ and\ \bibinfo {author} {\bibfnamefont {C.}~\bibnamefont
  {Barbieri}},\ }\href {\doibase 10.1016/j.ppnp.2004.02.038} {\bibfield
  {journal} {\bibinfo  {journal} {Prog. Part. Nucl. Phys.}\ }\textbf {\bibinfo
  {volume} {52}},\ \bibinfo {pages} {377} (\bibinfo {year} {2004})},\ \Eprint
  {http://arxiv.org/abs/nucl-th/0402034} {arXiv:nucl-th/0402034 [nucl-th]}
  \BibitemShut {NoStop}%
\bibitem [{\citenamefont {Barbieri}\ and\ \citenamefont
  {Dickhoff}(2003)}]{Barbieri2003DRPA}%
  \BibitemOpen
  \bibfield  {author} {\bibinfo {author} {\bibfnamefont {C.}~\bibnamefont
  {Barbieri}}\ and\ \bibinfo {author} {\bibfnamefont {W.~H.}\ \bibnamefont
  {Dickhoff}},\ }\href {\doibase 10.1103/PhysRevC.68.014311} {\bibfield
  {journal} {\bibinfo  {journal} {Phys. Rev. C}\ }\textbf {\bibinfo {volume}
  {68}},\ \bibinfo {pages} {014311} (\bibinfo {year} {2003})}\BibitemShut
  {NoStop}%
\bibitem [{\citenamefont {Barbieri}(2006)}]{Barbieri2006plbO16}%
  \BibitemOpen
  \bibfield  {author} {\bibinfo {author} {\bibfnamefont {C.}~\bibnamefont
  {Barbieri}},\ }\href@noop {} {\bibfield  {journal} {\bibinfo  {journal}
  {Phys. Lett. B}\ }\textbf {\bibinfo {volume} {643}},\ \bibinfo {pages} {268}
  (\bibinfo {year} {2006})}\BibitemShut {NoStop}%
\bibitem [{\citenamefont {Barbieri}\ \emph {et~al.}(2007)\citenamefont
  {Barbieri}, \citenamefont {Van~Neck},\ and\ \citenamefont
  {Dickhoff}}]{Barbieri:2007zy}%
  \BibitemOpen
  \bibfield  {author} {\bibinfo {author} {\bibfnamefont {C.}~\bibnamefont
  {Barbieri}}, \bibinfo {author} {\bibfnamefont {D.}~\bibnamefont {Van~Neck}},
  \ and\ \bibinfo {author} {\bibfnamefont {W.~H.}\ \bibnamefont {Dickhoff}},\
  }\href {\doibase 10.1103/PhysRevA.76.052503} {\bibfield  {journal} {\bibinfo
  {journal} {Phys. Rev.}\ }\textbf {\bibinfo {volume} {A76}},\ \bibinfo {pages}
  {052503} (\bibinfo {year} {2007})},\ \Eprint {http://arxiv.org/abs/0704.1542}
  {arXiv:0704.1542 [physics.chem-ph]} \BibitemShut {NoStop}%
\bibitem [{\citenamefont {Barbieri}\ and\ \citenamefont
  {Hjorth-Jensen}(2009)}]{Barbieri:2009nx}%
  \BibitemOpen
  \bibfield  {author} {\bibinfo {author} {\bibfnamefont {C.}~\bibnamefont
  {Barbieri}}\ and\ \bibinfo {author} {\bibfnamefont {M.}~\bibnamefont
  {Hjorth-Jensen}},\ }\href {\doibase 10.1103/PhysRevC.79.064313} {\bibfield
  {journal} {\bibinfo  {journal} {Phys. Rev.}\ }\textbf {\bibinfo {volume}
  {C79}},\ \bibinfo {pages} {064313} (\bibinfo {year} {2009})},\ \Eprint
  {http://arxiv.org/abs/0902.3942} {arXiv:0902.3942 [nucl-th]} \BibitemShut
  {NoStop}%
\bibitem [{\citenamefont {Cipollone}\ \emph {et~al.}(2013)\citenamefont
  {Cipollone}, \citenamefont {Barbieri},\ and\ \citenamefont
  {Navrátil}}]{Cipollone:2013zma}%
  \BibitemOpen
  \bibfield  {author} {\bibinfo {author} {\bibfnamefont {A.}~\bibnamefont
  {Cipollone}}, \bibinfo {author} {\bibfnamefont {C.}~\bibnamefont {Barbieri}},
  \ and\ \bibinfo {author} {\bibfnamefont {P.}~\bibnamefont {Navrátil}},\
  }\href {\doibase 10.1103/PhysRevLett.111.062501} {\bibfield  {journal}
  {\bibinfo  {journal} {Phys. Rev. Lett.}\ }\textbf {\bibinfo {volume} {111}},\
  \bibinfo {pages} {062501} (\bibinfo {year} {2013})},\ \Eprint
  {http://arxiv.org/abs/1303.4900} {arXiv:1303.4900 [nucl-th]} \BibitemShut
  {NoStop}%
\bibitem [{\citenamefont {Soma}\ \emph {et~al.}(2011)\citenamefont {Soma},
  \citenamefont {Duguet},\ and\ \citenamefont {Barbieri}}]{Soma:2011aj}%
  \BibitemOpen
  \bibfield  {author} {\bibinfo {author} {\bibfnamefont {V.}~\bibnamefont
  {Soma}}, \bibinfo {author} {\bibfnamefont {T.}~\bibnamefont {Duguet}}, \ and\
  \bibinfo {author} {\bibfnamefont {C.}~\bibnamefont {Barbieri}},\ }\href
  {\doibase 10.1103/PhysRevC.84.064317} {\bibfield  {journal} {\bibinfo
  {journal} {Phys. Rev.}\ }\textbf {\bibinfo {volume} {C84}},\ \bibinfo {pages}
  {064317} (\bibinfo {year} {2011})},\ \Eprint {http://arxiv.org/abs/1109.6230}
  {arXiv:1109.6230 [nucl-th]} \BibitemShut {NoStop}%
\bibitem [{\citenamefont {Soma}\ \emph {et~al.}(2013)\citenamefont {Soma},
  \citenamefont {Barbieri},\ and\ \citenamefont {Duguet}}]{Soma:2012zd}%
  \BibitemOpen
  \bibfield  {author} {\bibinfo {author} {\bibfnamefont {V.}~\bibnamefont
  {Soma}}, \bibinfo {author} {\bibfnamefont {C.}~\bibnamefont {Barbieri}}, \
  and\ \bibinfo {author} {\bibfnamefont {T.}~\bibnamefont {Duguet}},\ }\href
  {\doibase 10.1103/PhysRevC.87.011303} {\bibfield  {journal} {\bibinfo
  {journal} {Phys. Rev.}\ }\textbf {\bibinfo {volume} {C87}},\ \bibinfo {pages}
  {011303} (\bibinfo {year} {2013})},\ \Eprint {http://arxiv.org/abs/1208.2472}
  {arXiv:1208.2472 [nucl-th]} \BibitemShut {NoStop}%
\bibitem [{\citenamefont {Soma}\ \emph {et~al.}(2014)\citenamefont {Soma},
  \citenamefont {Barbieri},\ and\ \citenamefont {Duguet}}]{Soma:2013ona}%
  \BibitemOpen
  \bibfield  {author} {\bibinfo {author} {\bibfnamefont {V.}~\bibnamefont
  {Soma}}, \bibinfo {author} {\bibfnamefont {C.}~\bibnamefont {Barbieri}}, \
  and\ \bibinfo {author} {\bibfnamefont {T.}~\bibnamefont {Duguet}},\ }\href
  {\doibase 10.1103/PhysRevC.89.024323} {\bibfield  {journal} {\bibinfo
  {journal} {Phys. Rev.}\ }\textbf {\bibinfo {volume} {C89}},\ \bibinfo {pages}
  {024323} (\bibinfo {year} {2014})},\ \Eprint {http://arxiv.org/abs/1311.1989}
  {arXiv:1311.1989 [nucl-th]} \BibitemShut {NoStop}%
\bibitem [{\citenamefont {Carbone}\ \emph {et~al.}(2013)\citenamefont
  {Carbone}, \citenamefont {Cipollone}, \citenamefont {Barbieri}, \citenamefont
  {Rios},\ and\ \citenamefont {Polls}}]{Carbone:2013eqa}%
  \BibitemOpen
  \bibfield  {author} {\bibinfo {author} {\bibfnamefont {A.}~\bibnamefont
  {Carbone}}, \bibinfo {author} {\bibfnamefont {A.}~\bibnamefont {Cipollone}},
  \bibinfo {author} {\bibfnamefont {C.}~\bibnamefont {Barbieri}}, \bibinfo
  {author} {\bibfnamefont {A.}~\bibnamefont {Rios}}, \ and\ \bibinfo {author}
  {\bibfnamefont {A.}~\bibnamefont {Polls}},\ }\href {\doibase
  10.1103/PhysRevC.88.054326} {\bibfield  {journal} {\bibinfo  {journal} {Phys.
  Rev.}\ }\textbf {\bibinfo {volume} {C88}},\ \bibinfo {pages} {054326}
  (\bibinfo {year} {2013})},\ \Eprint {http://arxiv.org/abs/1310.3688}
  {arXiv:1310.3688 [nucl-th]} \BibitemShut {NoStop}%
\bibitem [{\citenamefont {Carbone}\ \emph {et~al.}(2014)\citenamefont
  {Carbone}, \citenamefont {Rios},\ and\ \citenamefont
  {Polls}}]{Carbone2014corr3nf}%
  \BibitemOpen
  \bibfield  {author} {\bibinfo {author} {\bibfnamefont {A.}~\bibnamefont
  {Carbone}}, \bibinfo {author} {\bibfnamefont {A.}~\bibnamefont {Rios}}, \
  and\ \bibinfo {author} {\bibfnamefont {A.}~\bibnamefont {Polls}},\ }\href
  {\doibase 10.1103/PhysRevC.90.054322} {\bibfield  {journal} {\bibinfo
  {journal} {Phys. Rev. C}\ }\textbf {\bibinfo {volume} {90}},\ \bibinfo
  {pages} {054322} (\bibinfo {year} {2014})}\BibitemShut {NoStop}%
\bibitem [{\citenamefont {Ekström}\ \emph {et~al.}(2015)\citenamefont
  {Ekström}, \citenamefont {Jansen}, \citenamefont {Wendt}, \citenamefont
  {Hagen}, \citenamefont {Papenbrock}, \citenamefont {Carlsson}, \citenamefont
  {Forssén}, \citenamefont {Hjorth-Jensen}, \citenamefont {Navrátil},\ and\
  \citenamefont {Nazarewicz}}]{Ekstrom:2015rta}%
  \BibitemOpen
  \bibfield  {author} {\bibinfo {author} {\bibfnamefont {A.}~\bibnamefont
  {Ekström}}, \bibinfo {author} {\bibfnamefont {G.~R.}\ \bibnamefont {Jansen}},
  \bibinfo {author} {\bibfnamefont {K.~A.}\ \bibnamefont {Wendt}}, \bibinfo
  {author} {\bibfnamefont {G.}~\bibnamefont {Hagen}}, \bibinfo {author}
  {\bibfnamefont {T.}~\bibnamefont {Papenbrock}}, \bibinfo {author}
  {\bibfnamefont {B.~D.}\ \bibnamefont {Carlsson}}, \bibinfo {author}
  {\bibfnamefont {C.}~\bibnamefont {Forssén}}, \bibinfo {author} {\bibfnamefont
  {M.}~\bibnamefont {Hjorth-Jensen}}, \bibinfo {author} {\bibfnamefont
  {P.}~\bibnamefont {Navrátil}}, \ and\ \bibinfo {author} {\bibfnamefont
  {W.}~\bibnamefont {Nazarewicz}},\ }\href {\doibase
  10.1103/PhysRevC.91.051301} {\bibfield  {journal} {\bibinfo  {journal} {Phys.
  Rev.}\ }\textbf {\bibinfo {volume} {C91}},\ \bibinfo {pages} {051301}
  (\bibinfo {year} {2015})},\ \Eprint {http://arxiv.org/abs/1502.04682}
  {arXiv:1502.04682 [nucl-th]} \BibitemShut {NoStop}%
\bibitem [{\citenamefont {Hagen}\ \emph {et~al.}(2009)\citenamefont {Hagen},
  \citenamefont {Papenbrock},\ and\ \citenamefont {Dean}}]{Hagen2009PRLcom}%
  \BibitemOpen
  \bibfield  {author} {\bibinfo {author} {\bibfnamefont {G.}~\bibnamefont
  {Hagen}}, \bibinfo {author} {\bibfnamefont {T.}~\bibnamefont {Papenbrock}}, \
  and\ \bibinfo {author} {\bibfnamefont {D.~J.}\ \bibnamefont {Dean}},\ }\href
  {\doibase 10.1103/PhysRevLett.103.062503} {\bibfield  {journal} {\bibinfo
  {journal} {Phys. Rev. Lett.}\ }\textbf {\bibinfo {volume} {103}},\ \bibinfo
  {pages} {062503} (\bibinfo {year} {2009})}\BibitemShut {NoStop}%
\bibitem [{\citenamefont {Benhar}\ \emph {et~al.}(2005)\citenamefont {Benhar},
  \citenamefont {Farina}, \citenamefont {Nakamura}, \citenamefont {Sakuda},\
  and\ \citenamefont {Seki}}]{Benhar:2005dj}%
  \BibitemOpen
  \bibfield  {author} {\bibinfo {author} {\bibfnamefont {O.}~\bibnamefont
  {Benhar}}, \bibinfo {author} {\bibfnamefont {N.}~\bibnamefont {Farina}},
  \bibinfo {author} {\bibfnamefont {H.}~\bibnamefont {Nakamura}}, \bibinfo
  {author} {\bibfnamefont {M.}~\bibnamefont {Sakuda}}, \ and\ \bibinfo {author}
  {\bibfnamefont {R.}~\bibnamefont {Seki}},\ }\href {\doibase
  10.1103/PhysRevD.72.053005} {\bibfield  {journal} {\bibinfo  {journal} {Phys.
  Rev.}\ }\textbf {\bibinfo {volume} {D72}},\ \bibinfo {pages} {053005}
  (\bibinfo {year} {2005})},\ \Eprint {http://arxiv.org/abs/hep-ph/0506116}
  {arXiv:hep-ph/0506116 [hep-ph]} \BibitemShut {NoStop}%
\bibitem [{\citenamefont {Benhar}\ \emph {et~al.}(2008)\citenamefont {Benhar},
  \citenamefont {Day},\ and\ \citenamefont {Sick}}]{Benhar:2006wy}%
  \BibitemOpen
  \bibfield  {author} {\bibinfo {author} {\bibfnamefont {O.}~\bibnamefont
  {Benhar}}, \bibinfo {author} {\bibfnamefont {D.}~\bibnamefont {Day}}, \ and\
  \bibinfo {author} {\bibfnamefont {I.}~\bibnamefont {Sick}},\ }\href {\doibase
  10.1103/RevModPhys.80.189} {\bibfield  {journal} {\bibinfo  {journal} {Rev.
  Mod. Phys.}\ }\textbf {\bibinfo {volume} {80}},\ \bibinfo {pages} {189}
  (\bibinfo {year} {2008})},\ \Eprint {http://arxiv.org/abs/nucl-ex/0603029}
  {arXiv:nucl-ex/0603029 [nucl-ex]} \BibitemShut {NoStop}%
\bibitem [{\citenamefont {Ankowski}\ \emph {et~al.}(2015)\citenamefont
  {Ankowski}, \citenamefont {Benhar},\ and\ \citenamefont
  {Sakuda}}]{Ankowski:2014yfa}%
  \BibitemOpen
  \bibfield  {author} {\bibinfo {author} {\bibfnamefont {A.~M.}\ \bibnamefont
  {Ankowski}}, \bibinfo {author} {\bibfnamefont {O.}~\bibnamefont {Benhar}}, \
  and\ \bibinfo {author} {\bibfnamefont {M.}~\bibnamefont {Sakuda}},\ }\href
  {\doibase 10.1103/PhysRevD.91.033005} {\bibfield  {journal} {\bibinfo
  {journal} {Phys. Rev.}\ }\textbf {\bibinfo {volume} {D91}},\ \bibinfo {pages}
  {033005} (\bibinfo {year} {2015})},\ \Eprint {http://arxiv.org/abs/1404.5687}
  {arXiv:1404.5687 [nucl-th]} \BibitemShut {NoStop}%
\bibitem [{\citenamefont {Benhar}\ \emph {et~al.}(1991)\citenamefont {Benhar},
  \citenamefont {Fabrocini}, \citenamefont {Fantoni}, \citenamefont {Miller},
  \citenamefont {Pandharipande},\ and\ \citenamefont {Sick}}]{Benhar:1991af}%
  \BibitemOpen
  \bibfield  {author} {\bibinfo {author} {\bibfnamefont {O.}~\bibnamefont
  {Benhar}}, \bibinfo {author} {\bibfnamefont {A.}~\bibnamefont {Fabrocini}},
  \bibinfo {author} {\bibfnamefont {S.}~\bibnamefont {Fantoni}}, \bibinfo
  {author} {\bibfnamefont {G.~A.}\ \bibnamefont {Miller}}, \bibinfo {author}
  {\bibfnamefont {V.~R.}\ \bibnamefont {Pandharipande}}, \ and\ \bibinfo
  {author} {\bibfnamefont {I.}~\bibnamefont {Sick}},\ }\href {\doibase
  10.1103/PhysRevC.44.2328} {\bibfield  {journal} {\bibinfo  {journal} {Phys.
  Rev.}\ }\textbf {\bibinfo {volume} {C44}},\ \bibinfo {pages} {2328} (\bibinfo
  {year} {1991})}\BibitemShut {NoStop}%
\bibitem [{\citenamefont {Dickhoff}\ and\ \citenamefont
  {Van~Neck}(2008)}]{Dickhoff2008book}%
  \BibitemOpen
  \bibfield  {author} {\bibinfo {author} {\bibfnamefont {W.~H.}\ \bibnamefont
  {Dickhoff}}\ and\ \bibinfo {author} {\bibfnamefont {D.}~\bibnamefont
  {Van~Neck}},\ }\href@noop {} {\emph {\bibinfo {title} {Many-body theory
  exposed!}}},\ \bibinfo {edition} {2nd}\ ed.\ (\bibinfo  {publisher} {World
  Scientific Publishing, London},\ \bibinfo {year} {2008})\BibitemShut
  {NoStop}%
\bibitem [{\citenamefont {Schirmer}\ \emph {et~al.}(1983)\citenamefont
  {Schirmer}, \citenamefont {Cederbaum},\ and\ \citenamefont
  {Walter}}]{Schirmer1983}%
  \BibitemOpen
  \bibfield  {author} {\bibinfo {author} {\bibfnamefont {J.}~\bibnamefont
  {Schirmer}}, \bibinfo {author} {\bibfnamefont {L.~S.}\ \bibnamefont
  {Cederbaum}}, \ and\ \bibinfo {author} {\bibfnamefont {O.}~\bibnamefont
  {Walter}},\ }\href {\doibase 10.1103/PhysRevA.28.1237} {\bibfield  {journal}
  {\bibinfo  {journal} {Phys. Rev. A}\ }\textbf {\bibinfo {volume} {28}},\
  \bibinfo {pages} {1237} (\bibinfo {year} {1983})}\BibitemShut {NoStop}%
\bibitem [{\citenamefont {Barbieri}\ and\ \citenamefont
  {Carbone}(2017)}]{Barbieri2017LNP}%
  \BibitemOpen
  \bibfield  {author} {\bibinfo {author} {\bibfnamefont {C.}~\bibnamefont
  {Barbieri}}\ and\ \bibinfo {author} {\bibfnamefont {A.}~\bibnamefont
  {Carbone}},\ }\enquote {\bibinfo {title} {Self-consistent green's function
  approaches},}\ in\ \href@noop {} {\emph {\bibinfo {booktitle} {An Advanced
  Course in Computational Nuclear Physics: Bridging the Scales from Quarks to
  Neutron Stars}}},\ \bibinfo {series} {Lect. Notes Phys.}, Vol.\ \bibinfo
  {volume} {936},\ \bibinfo {editor} {edited by\ \bibinfo {editor}
  {\bibfnamefont {M.}~\bibnamefont {Hjorth-Jensen}}, \bibinfo {editor}
  {\bibfnamefont {M.~P.}\ \bibnamefont {Lombardo}}, \ and\ \bibinfo {editor}
  {\bibfnamefont {U.}~\bibnamefont {van Kolck}}}\ (\bibinfo  {publisher}
  {Springer International Publishing},\ \bibinfo {address} {Cham},\ \bibinfo
  {year} {2017})\ Chap.~\bibinfo {chapter} {11}, pp.\ \bibinfo {pages}
  {571--644}\BibitemShut {NoStop}%
\bibitem [{\citenamefont {Hagen}\ \emph {et~al.}(2007)\citenamefont {Hagen},
  \citenamefont {Papenbrock}, \citenamefont {Dean}, \citenamefont {Schwenk},
  \citenamefont {Nogga}, \citenamefont {Wloch},\ and\ \citenamefont
  {Piecuch}}]{Hagen:2007ew}%
  \BibitemOpen
  \bibfield  {author} {\bibinfo {author} {\bibfnamefont {G.}~\bibnamefont
  {Hagen}}, \bibinfo {author} {\bibfnamefont {T.}~\bibnamefont {Papenbrock}},
  \bibinfo {author} {\bibfnamefont {D.~J.}\ \bibnamefont {Dean}}, \bibinfo
  {author} {\bibfnamefont {A.}~\bibnamefont {Schwenk}}, \bibinfo {author}
  {\bibfnamefont {A.}~\bibnamefont {Nogga}}, \bibinfo {author} {\bibfnamefont
  {M.}~\bibnamefont {Wloch}}, \ and\ \bibinfo {author} {\bibfnamefont
  {P.}~\bibnamefont {Piecuch}},\ }\href {\doibase 10.1103/PhysRevC.76.034302}
  {\bibfield  {journal} {\bibinfo  {journal} {Phys. Rev.}\ }\textbf {\bibinfo
  {volume} {C76}},\ \bibinfo {pages} {034302} (\bibinfo {year} {2007})},\
  \Eprint {http://arxiv.org/abs/0704.2854} {arXiv:0704.2854 [nucl-th]}
  \BibitemShut {NoStop}%
\bibitem [{\citenamefont {Roth}\ \emph {et~al.}(2012)\citenamefont {Roth},
  \citenamefont {Binder}, \citenamefont {Vobig}, \citenamefont {Calci},
  \citenamefont {Langhammer},\ and\ \citenamefont {Navratil}}]{Roth:2011vt}%
  \BibitemOpen
  \bibfield  {author} {\bibinfo {author} {\bibfnamefont {R.}~\bibnamefont
  {Roth}}, \bibinfo {author} {\bibfnamefont {S.}~\bibnamefont {Binder}},
  \bibinfo {author} {\bibfnamefont {K.}~\bibnamefont {Vobig}}, \bibinfo
  {author} {\bibfnamefont {A.}~\bibnamefont {Calci}}, \bibinfo {author}
  {\bibfnamefont {J.}~\bibnamefont {Langhammer}}, \ and\ \bibinfo {author}
  {\bibfnamefont {P.}~\bibnamefont {Navratil}},\ }\href {\doibase
  10.1103/PhysRevLett.109.052501} {\bibfield  {journal} {\bibinfo  {journal}
  {Phys. Rev. Lett.}\ }\textbf {\bibinfo {volume} {109}},\ \bibinfo {pages}
  {052501} (\bibinfo {year} {2012})},\ \Eprint {http://arxiv.org/abs/1112.0287}
  {arXiv:1112.0287 [nucl-th]} \BibitemShut {NoStop}%
\bibitem [{\citenamefont {Barbieri}(2014)}]{Barbieri2014QMBT}%
  \BibitemOpen
  \bibfield  {author} {\bibinfo {author} {\bibfnamefont {C.}~\bibnamefont
  {Barbieri}},\ }\href {http://stacks.iop.org/1742-6596/529/i=1/a=012005}
  {\bibfield  {journal} {\bibinfo  {journal} {Journal of Physics: Conference
  Series}\ }\textbf {\bibinfo {volume} {529}},\ \bibinfo {pages} {012005}
  (\bibinfo {year} {2014})}\BibitemShut {NoStop}%
\bibitem [{\citenamefont {Cipollone}\ \emph {et~al.}(2015)\citenamefont
  {Cipollone}, \citenamefont {Barbieri},\ and\ \citenamefont
  {Navrátil}}]{Cipollone:2014hfa}%
  \BibitemOpen
  \bibfield  {author} {\bibinfo {author} {\bibfnamefont {A.}~\bibnamefont
  {Cipollone}}, \bibinfo {author} {\bibfnamefont {C.}~\bibnamefont {Barbieri}},
  \ and\ \bibinfo {author} {\bibfnamefont {P.}~\bibnamefont {Navrátil}},\
  }\href {\doibase 10.1103/PhysRevC.92.014306} {\bibfield  {journal} {\bibinfo
  {journal} {Phys. Rev.}\ }\textbf {\bibinfo {volume} {C92}},\ \bibinfo {pages}
  {014306} (\bibinfo {year} {2015})},\ \Eprint {http://arxiv.org/abs/1412.0491}
  {arXiv:1412.0491 [nucl-th]} \BibitemShut {NoStop}%
\bibitem [{\citenamefont {Raimondi}\ and\ \citenamefont
  {Barbieri}(2017)}]{Raimondi:2017kzi}%
  \BibitemOpen
  \bibfield  {author} {\bibinfo {author} {\bibfnamefont {F.}~\bibnamefont
  {Raimondi}}\ and\ \bibinfo {author} {\bibfnamefont {C.}~\bibnamefont
  {Barbieri}},\ }\href@noop {} {\  (\bibinfo {year} {2017})},\ \Eprint
  {http://arxiv.org/abs/1709.04330} {arXiv:1709.04330 [nucl-th]} \BibitemShut
  {NoStop}%
\bibitem [{\citenamefont {Rios}\ \emph {et~al.}(2017)\citenamefont {Rios},
  \citenamefont {Carbone},\ and\ \citenamefont {Polls}}]{Rios2017SumRules}%
  \BibitemOpen
  \bibfield  {author} {\bibinfo {author} {\bibfnamefont {A.}~\bibnamefont
  {Rios}}, \bibinfo {author} {\bibfnamefont {A.}~\bibnamefont {Carbone}}, \
  and\ \bibinfo {author} {\bibfnamefont {A.}~\bibnamefont {Polls}},\ }\href
  {\doibase 10.1103/PhysRevC.96.014003} {\bibfield  {journal} {\bibinfo
  {journal} {Phys. Rev. C}\ }\textbf {\bibinfo {volume} {96}},\ \bibinfo
  {pages} {014003} (\bibinfo {year} {2017})}\BibitemShut {NoStop}%
\bibitem [{\citenamefont {Carlson}\ \emph {et~al.}(2015)\citenamefont
  {Carlson}, \citenamefont {Gandolfi}, \citenamefont {Pederiva}, \citenamefont
  {Pieper}, \citenamefont {Schiavilla}, \citenamefont {Schmidt},\ and\
  \citenamefont {Wiringa}}]{Carlson:2014vla}%
  \BibitemOpen
  \bibfield  {author} {\bibinfo {author} {\bibfnamefont {J.}~\bibnamefont
  {Carlson}}, \bibinfo {author} {\bibfnamefont {S.}~\bibnamefont {Gandolfi}},
  \bibinfo {author} {\bibfnamefont {F.}~\bibnamefont {Pederiva}}, \bibinfo
  {author} {\bibfnamefont {S.~C.}\ \bibnamefont {Pieper}}, \bibinfo {author}
  {\bibfnamefont {R.}~\bibnamefont {Schiavilla}}, \bibinfo {author}
  {\bibfnamefont {K.~E.}\ \bibnamefont {Schmidt}}, \ and\ \bibinfo {author}
  {\bibfnamefont {R.~B.}\ \bibnamefont {Wiringa}},\ }\href {\doibase
  10.1103/RevModPhys.87.1067} {\bibfield  {journal} {\bibinfo  {journal} {Rev.
  Mod. Phys.}\ }\textbf {\bibinfo {volume} {87}},\ \bibinfo {pages} {1067}
  (\bibinfo {year} {2015})},\ \Eprint {http://arxiv.org/abs/1412.3081}
  {arXiv:1412.3081 [nucl-th]} \BibitemShut {NoStop}%
\bibitem [{\citenamefont {Metropolis}\ \emph {et~al.}(1953)\citenamefont
  {Metropolis}, \citenamefont {Rosenbluth}, \citenamefont {Rosenbluth},
  \citenamefont {Teller},\ and\ \citenamefont {Teller}}]{Metropolis:1953am}%
  \BibitemOpen
  \bibfield  {author} {\bibinfo {author} {\bibfnamefont {N.}~\bibnamefont
  {Metropolis}}, \bibinfo {author} {\bibfnamefont {A.~W.}\ \bibnamefont
  {Rosenbluth}}, \bibinfo {author} {\bibfnamefont {M.~N.}\ \bibnamefont
  {Rosenbluth}}, \bibinfo {author} {\bibfnamefont {A.~H.}\ \bibnamefont
  {Teller}}, \ and\ \bibinfo {author} {\bibfnamefont {E.}~\bibnamefont
  {Teller}},\ }\href {\doibase 10.1063/1.1699114} {\bibfield  {journal}
  {\bibinfo  {journal} {J. Chem. Phys.}\ }\textbf {\bibinfo {volume} {21}},\
  \bibinfo {pages} {1087} (\bibinfo {year} {1953})}\BibitemShut {NoStop}%
\bibitem [{\citenamefont {Schmidt}\ and\ \citenamefont
  {Fantoni}(1999)}]{Schmidt:1999lik}%
  \BibitemOpen
  \bibfield  {author} {\bibinfo {author} {\bibfnamefont {K.~E.}\ \bibnamefont
  {Schmidt}}\ and\ \bibinfo {author} {\bibfnamefont {S.}~\bibnamefont
  {Fantoni}},\ }\href {\doibase 10.1016/S0370-2693(98)01522-6} {\bibfield
  {journal} {\bibinfo  {journal} {Phys. Lett.}\ }\textbf {\bibinfo {volume}
  {B446}},\ \bibinfo {pages} {99} (\bibinfo {year} {1999})}\BibitemShut
  {NoStop}%
\bibitem [{\citenamefont {Pieper}\ \emph {et~al.}(1990)\citenamefont {Pieper},
  \citenamefont {Wiringa},\ and\ \citenamefont
  {Pandharipande}}]{Pieper:1990ji}%
  \BibitemOpen
  \bibfield  {author} {\bibinfo {author} {\bibfnamefont {S.~C.}\ \bibnamefont
  {Pieper}}, \bibinfo {author} {\bibfnamefont {R.~B.}\ \bibnamefont {Wiringa}},
  \ and\ \bibinfo {author} {\bibfnamefont {V.~R.}\ \bibnamefont
  {Pandharipande}},\ }\href {\doibase 10.1103/PhysRevLett.64.364} {\bibfield
  {journal} {\bibinfo  {journal} {Phys. Rev. Lett.}\ }\textbf {\bibinfo
  {volume} {64}},\ \bibinfo {pages} {364} (\bibinfo {year} {1990})}\BibitemShut
  {NoStop}%
\bibitem [{\citenamefont {Brida}\ \emph {et~al.}(2011)\citenamefont {Brida},
  \citenamefont {Pieper},\ and\ \citenamefont {Wiringa}}]{Brida:2011yp}%
  \BibitemOpen
  \bibfield  {author} {\bibinfo {author} {\bibfnamefont {I.}~\bibnamefont
  {Brida}}, \bibinfo {author} {\bibfnamefont {S.~C.}\ \bibnamefont {Pieper}}, \
  and\ \bibinfo {author} {\bibfnamefont {R.~B.}\ \bibnamefont {Wiringa}},\
  }\href {\doibase 10.1103/PhysRevC.84.024319} {\bibfield  {journal} {\bibinfo
  {journal} {Phys. Rev.}\ }\textbf {\bibinfo {volume} {C84}},\ \bibinfo {pages}
  {024319} (\bibinfo {year} {2011})},\ \Eprint {http://arxiv.org/abs/1106.3121}
  {arXiv:1106.3121 [nucl-th]} \BibitemShut {NoStop}%
\bibitem [{\citenamefont {Pastore}\ \emph {et~al.}(2013)\citenamefont
  {Pastore}, \citenamefont {Pieper}, \citenamefont {Schiavilla},\ and\
  \citenamefont {Wiringa}}]{Pastore:2012rp}%
  \BibitemOpen
  \bibfield  {author} {\bibinfo {author} {\bibfnamefont {S.}~\bibnamefont
  {Pastore}}, \bibinfo {author} {\bibfnamefont {S.~C.}\ \bibnamefont {Pieper}},
  \bibinfo {author} {\bibfnamefont {R.}~\bibnamefont {Schiavilla}}, \ and\
  \bibinfo {author} {\bibfnamefont {R.~B.}\ \bibnamefont {Wiringa}},\ }\href
  {\doibase 10.1103/PhysRevC.87.035503} {\bibfield  {journal} {\bibinfo
  {journal} {Phys. Rev.}\ }\textbf {\bibinfo {volume} {C87}},\ \bibinfo {pages}
  {035503} (\bibinfo {year} {2013})},\ \Eprint {http://arxiv.org/abs/1212.3375}
  {arXiv:1212.3375 [nucl-th]} \BibitemShut {NoStop}%
\bibitem [{\citenamefont {Benhar}\ \emph {et~al.}(2017)\citenamefont {Benhar},
  \citenamefont {Huber}, \citenamefont {Mariani},\ and\ \citenamefont
  {Meloni}}]{Benhar:2015wva}%
  \BibitemOpen
  \bibfield  {author} {\bibinfo {author} {\bibfnamefont {O.}~\bibnamefont
  {Benhar}}, \bibinfo {author} {\bibfnamefont {P.}~\bibnamefont {Huber}},
  \bibinfo {author} {\bibfnamefont {C.}~\bibnamefont {Mariani}}, \ and\
  \bibinfo {author} {\bibfnamefont {D.}~\bibnamefont {Meloni}},\ }\href
  {\doibase 10.1016/j.physrep.2017.07.004} {\bibfield  {journal} {\bibinfo
  {journal} {Phys. Rept.}\ }\textbf {\bibinfo {volume} {700}},\ \bibinfo
  {pages} {1} (\bibinfo {year} {2017})},\ \Eprint
  {http://arxiv.org/abs/1501.06448} {arXiv:1501.06448 [nucl-th]} \BibitemShut
  {NoStop}%
\bibitem [{\citenamefont {Cooper}\ \emph {et~al.}(1993)\citenamefont {Cooper},
  \citenamefont {Hama}, \citenamefont {Clark},\ and\ \citenamefont
  {Mercer}}]{Cooper:1993nx}%
  \BibitemOpen
  \bibfield  {author} {\bibinfo {author} {\bibfnamefont {E.~D.}\ \bibnamefont
  {Cooper}}, \bibinfo {author} {\bibfnamefont {S.}~\bibnamefont {Hama}},
  \bibinfo {author} {\bibfnamefont {B.~C.}\ \bibnamefont {Clark}}, \ and\
  \bibinfo {author} {\bibfnamefont {R.~L.}\ \bibnamefont {Mercer}},\ }\href
  {\doibase 10.1103/PhysRevC.47.297} {\bibfield  {journal} {\bibinfo  {journal}
  {Phys. Rev.}\ }\textbf {\bibinfo {volume} {C47}},\ \bibinfo {pages} {297}
  (\bibinfo {year} {1993})}\BibitemShut {NoStop}%
\bibitem [{\citenamefont {Garcia~Ruiz}\ \emph {et~al.}(2016)\citenamefont
  {Garcia~Ruiz}, \citenamefont {Bissell}, \citenamefont {Blaum}, \citenamefont
  {Ekstr{\"o}m}, \citenamefont {Fr{\"o}mmgen}, \citenamefont {Hagen},
  \citenamefont {Hammen}, \citenamefont {Hebeler}, \citenamefont {Holt},
  \citenamefont {Jansen}, \citenamefont {Kowalska}, \citenamefont {Kreim},
  \citenamefont {Nazarewicz}, \citenamefont {Neugart}, \citenamefont {Neyens},
  \citenamefont {N{\"o}rtersh{\"a}user}, \citenamefont {Papenbrock},
  \citenamefont {Papuga}, \citenamefont {Schwenk}, \citenamefont {Simonis},
  \citenamefont {Wendt},\ and\ \citenamefont {Yordanov}}]{Ruiz2015Nature}%
  \BibitemOpen
  \bibfield  {author} {\bibinfo {author} {\bibfnamefont {R.~F.}\ \bibnamefont
  {Garcia~Ruiz}}, \bibinfo {author} {\bibfnamefont {M.~L.}\ \bibnamefont
  {Bissell}}, \bibinfo {author} {\bibfnamefont {K.}~\bibnamefont {Blaum}},
  \bibinfo {author} {\bibfnamefont {A.}~\bibnamefont {Ekstr{\"o}m}}, \bibinfo
  {author} {\bibfnamefont {N.}~\bibnamefont {Fr{\"o}mmgen}}, \bibinfo {author}
  {\bibfnamefont {G.}~\bibnamefont {Hagen}}, \bibinfo {author} {\bibfnamefont
  {M.}~\bibnamefont {Hammen}}, \bibinfo {author} {\bibfnamefont
  {K.}~\bibnamefont {Hebeler}}, \bibinfo {author} {\bibfnamefont {J.~D.}\
  \bibnamefont {Holt}}, \bibinfo {author} {\bibfnamefont {G.~R.}\ \bibnamefont
  {Jansen}}, \bibinfo {author} {\bibfnamefont {M.}~\bibnamefont {Kowalska}},
  \bibinfo {author} {\bibfnamefont {K.}~\bibnamefont {Kreim}}, \bibinfo
  {author} {\bibfnamefont {W.}~\bibnamefont {Nazarewicz}}, \bibinfo {author}
  {\bibfnamefont {R.}~\bibnamefont {Neugart}}, \bibinfo {author} {\bibfnamefont
  {G.}~\bibnamefont {Neyens}}, \bibinfo {author} {\bibfnamefont
  {W.}~\bibnamefont {N{\"o}rtersh{\"a}user}}, \bibinfo {author} {\bibfnamefont
  {T.}~\bibnamefont {Papenbrock}}, \bibinfo {author} {\bibfnamefont
  {J.}~\bibnamefont {Papuga}}, \bibinfo {author} {\bibfnamefont
  {A.}~\bibnamefont {Schwenk}}, \bibinfo {author} {\bibfnamefont
  {J.}~\bibnamefont {Simonis}}, \bibinfo {author} {\bibfnamefont {K.~A.}\
  \bibnamefont {Wendt}}, \ and\ \bibinfo {author} {\bibfnamefont {D.~T.}\
  \bibnamefont {Yordanov}},\ }\href@noop {} {\bibfield  {journal} {\bibinfo
  {journal} {Nature Physics}\ }\textbf {\bibinfo {volume} {12}},\ \bibinfo
  {pages} {594 EP } (\bibinfo {year} {2016})}\BibitemShut {NoStop}%
\bibitem [{\citenamefont {Lapoux}\ \emph {et~al.}(2016)\citenamefont {Lapoux},
  \citenamefont {Somà}, \citenamefont {Barbieri}, \citenamefont {Hergert},
  \citenamefont {Holt},\ and\ \citenamefont {Stroberg}}]{Lapoux:2016exf}%
  \BibitemOpen
  \bibfield  {author} {\bibinfo {author} {\bibfnamefont {V.}~\bibnamefont
  {Lapoux}}, \bibinfo {author} {\bibfnamefont {V.}~\bibnamefont {Somà}},
  \bibinfo {author} {\bibfnamefont {C.}~\bibnamefont {Barbieri}}, \bibinfo
  {author} {\bibfnamefont {H.}~\bibnamefont {Hergert}}, \bibinfo {author}
  {\bibfnamefont {J.~D.}\ \bibnamefont {Holt}}, \ and\ \bibinfo {author}
  {\bibfnamefont {S.}~\bibnamefont {Stroberg}},\ }\href {\doibase
  10.1103/PhysRevLett.117.052501} {\bibfield  {journal} {\bibinfo  {journal}
  {Phys. Rev. Lett.}\ }\textbf {\bibinfo {volume} {117}},\ \bibinfo {pages}
  {052501} (\bibinfo {year} {2016})},\ \Eprint
  {http://arxiv.org/abs/1605.07885} {arXiv:1605.07885 [nucl-ex]} \BibitemShut
  {NoStop}%
\bibitem [{\citenamefont {De~Vries}\ \emph {et~al.}(1987)\citenamefont
  {De~Vries}, \citenamefont {De~Jager},\ and\ \citenamefont
  {De~Vries}}]{DeJager:1987qc}%
  \BibitemOpen
  \bibfield  {author} {\bibinfo {author} {\bibfnamefont {H.}~\bibnamefont
  {De~Vries}}, \bibinfo {author} {\bibfnamefont {C.~W.}\ \bibnamefont
  {De~Jager}}, \ and\ \bibinfo {author} {\bibfnamefont {C.}~\bibnamefont
  {De~Vries}},\ }\href {\doibase 10.1016/0092-640X(87)90013-1} {\bibfield
  {journal} {\bibinfo  {journal} {Atom. Data Nucl. Data Tabl.}\ }\textbf
  {\bibinfo {volume} {36}},\ \bibinfo {pages} {495} (\bibinfo {year}
  {1987})}\BibitemShut {NoStop}%
\bibitem [{\citenamefont {Lonardoni}\ \emph {et~al.}(2017)\citenamefont
  {Lonardoni}, \citenamefont {Lovato}, \citenamefont {Pieper},\ and\
  \citenamefont {Wiringa}}]{Lonardoni:2017egu}%
  \BibitemOpen
  \bibfield  {author} {\bibinfo {author} {\bibfnamefont {D.}~\bibnamefont
  {Lonardoni}}, \bibinfo {author} {\bibfnamefont {A.}~\bibnamefont {Lovato}},
  \bibinfo {author} {\bibfnamefont {S.~C.}\ \bibnamefont {Pieper}}, \ and\
  \bibinfo {author} {\bibfnamefont {R.~B.}\ \bibnamefont {Wiringa}},\ }\href
  {\doibase 10.1103/PhysRevC.96.024326} {\bibfield  {journal} {\bibinfo
  {journal} {Phys. Rev.}\ }\textbf {\bibinfo {volume} {C96}},\ \bibinfo {pages}
  {024326} (\bibinfo {year} {2017})},\ \Eprint
  {http://arxiv.org/abs/1705.04337} {arXiv:1705.04337 [nucl-th]} \BibitemShut
  {NoStop}%
\bibitem [{\citenamefont {Marcucci}\ \emph {et~al.}(2016)\citenamefont
  {Marcucci}, \citenamefont {Gross}, \citenamefont {Pena}, \citenamefont
  {Piarulli}, \citenamefont {Schiavilla}, \citenamefont {Sick}, \citenamefont
  {Stadler}, \citenamefont {Van~Orden},\ and\ \citenamefont
  {Viviani}}]{Marcucci:2015rca}%
  \BibitemOpen
  \bibfield  {author} {\bibinfo {author} {\bibfnamefont {L.~E.}\ \bibnamefont
  {Marcucci}}, \bibinfo {author} {\bibfnamefont {F.}~\bibnamefont {Gross}},
  \bibinfo {author} {\bibfnamefont {M.~T.}\ \bibnamefont {Pena}}, \bibinfo
  {author} {\bibfnamefont {M.}~\bibnamefont {Piarulli}}, \bibinfo {author}
  {\bibfnamefont {R.}~\bibnamefont {Schiavilla}}, \bibinfo {author}
  {\bibfnamefont {I.}~\bibnamefont {Sick}}, \bibinfo {author} {\bibfnamefont
  {A.}~\bibnamefont {Stadler}}, \bibinfo {author} {\bibfnamefont {J.~W.}\
  \bibnamefont {Van~Orden}}, \ and\ \bibinfo {author} {\bibfnamefont
  {M.}~\bibnamefont {Viviani}},\ }\href {\doibase
  10.1088/0954-3899/43/2/023002} {\bibfield  {journal} {\bibinfo  {journal} {J.
  Phys.}\ }\textbf {\bibinfo {volume} {G43}},\ \bibinfo {pages} {023002}
  (\bibinfo {year} {2016})},\ \Eprint {http://arxiv.org/abs/1504.05063}
  {arXiv:1504.05063 [nucl-th]} \BibitemShut {NoStop}%
\bibitem [{\citenamefont {Lonardoni}\ \emph {et~al.}(2018)\citenamefont
  {Lonardoni}, \citenamefont {Gandolfi}, \citenamefont {Lynn}, \citenamefont
  {Petrie}, \citenamefont {Carlson}, \citenamefont {Schmidt},\ and\
  \citenamefont {Schwenk}}]{Lonardoni:2018nob}%
  \BibitemOpen
  \bibfield  {author} {\bibinfo {author} {\bibfnamefont {D.}~\bibnamefont
  {Lonardoni}}, \bibinfo {author} {\bibfnamefont {S.}~\bibnamefont {Gandolfi}},
  \bibinfo {author} {\bibfnamefont {J.~E.}\ \bibnamefont {Lynn}}, \bibinfo
  {author} {\bibfnamefont {C.}~\bibnamefont {Petrie}}, \bibinfo {author}
  {\bibfnamefont {J.}~\bibnamefont {Carlson}}, \bibinfo {author} {\bibfnamefont
  {K.~E.}\ \bibnamefont {Schmidt}}, \ and\ \bibinfo {author} {\bibfnamefont
  {A.}~\bibnamefont {Schwenk}},\ }\href@noop {} {\  (\bibinfo {year} {2018})},\
  \Eprint {http://arxiv.org/abs/1802.08932} {arXiv:1802.08932 [nucl-th]}
  \BibitemShut {NoStop}%
\bibitem [{\citenamefont {Benhar}\ \emph {et~al.}(2006)\citenamefont {Benhar},
  \citenamefont {Day},\ and\ \citenamefont {Sick}}]{Benhar:2006er}%
  \BibitemOpen
  \bibfield  {author} {\bibinfo {author} {\bibfnamefont {O.}~\bibnamefont
  {Benhar}}, \bibinfo {author} {\bibfnamefont {D.}~\bibnamefont {Day}}, \ and\
  \bibinfo {author} {\bibfnamefont {I.}~\bibnamefont {Sick}},\ }\href@noop {}
  {\  (\bibinfo {year} {2006})},\ \Eprint
  {http://arxiv.org/abs/nucl-ex/0603032} {arXiv:nucl-ex/0603032 [nucl-ex]}
  \BibitemShut {NoStop}%
\bibitem [{\citenamefont {Som\`a}\ \emph {et~al.}(2014)\citenamefont {Som\`a},
  \citenamefont {Cipollone}, \citenamefont {Barbieri}, \citenamefont
  {Navrátil},\ and\ \citenamefont {Duguet}}]{Soma:2013xha}%
  \BibitemOpen
  \bibfield  {author} {\bibinfo {author} {\bibfnamefont {V.}~\bibnamefont
  {Som\`a}}, \bibinfo {author} {\bibfnamefont {A.}~\bibnamefont {Cipollone}},
  \bibinfo {author} {\bibfnamefont {C.}~\bibnamefont {Barbieri}}, \bibinfo
  {author} {\bibfnamefont {P.}~\bibnamefont {Navrátil}}, \ and\ \bibinfo
  {author} {\bibfnamefont {T.}~\bibnamefont {Duguet}},\ }\href {\doibase
  10.1103/PhysRevC.89.061301} {\bibfield  {journal} {\bibinfo  {journal} {Phys.
  Rev.}\ }\textbf {\bibinfo {volume} {C89}},\ \bibinfo {pages} {061301}
  (\bibinfo {year} {2014})},\ \Eprint {http://arxiv.org/abs/1312.2068}
  {arXiv:1312.2068 [nucl-th]} \BibitemShut {NoStop}%
\end{thebibliography}%
\end{document}